\documentclass[manuscript]{acmart}

\AtBeginDocument{%
  }

\setcopyright{acmlicensed}
\copyrightyear{2027}
\acmYear{2027}
\acmDOI{XXXXXXX.XXXXXXX}

\acmConference[CHI '27]{ACM Conference on Human Factors in Computing Systems}{May 10--14, 2027}{Pittsburgh, USA}

\acmISBN{978-1-4503-XXXX-X/2018/06}

\usepackage{url}
\usepackage{hyperref}

\begin{document}

\title{Altruism as Infrastructure: Volunteer Moderation in a Bangladeshi Higher Education Facebook Group}

\author{Umme Jannat Taposhi\textsuperscript{*}}
\affiliation{%
  \institution{BRAC University}
  \city{Dhaka}
  \country{Bangladesh}
}

\author{Md. Tawhid Anwar\textsuperscript{*}}
\affiliation{%
  \institution{BRAC University}
  \city{Dhaka}
  \country{Bangladesh}
}

\author{Alvi Islam Ratul}
\affiliation{%
  \institution{BRAC University}
  \city{Dhaka}
  \country{Bangladesh}
}

\author{S M Taiabul Haque\textsuperscript{\dag}}
\affiliation{%
  \institution{BRAC University}
  \city{Dhaka}
  \country{Bangladesh}
}

\renewcommand{\shortauthors}{Taposhi et al.}

\begin{abstract}
Aspiring international students across Asian countries increasingly depend on commercial education agents to navigate scholarships, documentation, and visas. Alongside this commercial infrastructure, volunteer-run Facebook groups have emerged. Unpaid admins and moderators, often under their real identities, vet information, screen scams, and guide members through scholarships, visas, and departure logistics. We study one such Bangladeshi group, \textit{HigherStudyAbroad: Global Hub of Bangladeshis}, founded in 2010. Drawing on semi-structured interviews with 17 volunteer admins and moderators, we found that altruism becomes an organizing logic. It shapes moderators' identities, sustains invisible labor, and grants the group legitimacy against paid agents, even amid new technologies such as AI that could reshape this work. Human judgment remains central to how moderators sustain trust. We discuss implications for HCI's understanding of volunteer labor and migration infrastructure, along with design implications for platforms that depend on unpaid, long-term contribution. 
\end{abstract}


\maketitle
\section{Introduction}
International student mobility has roughly tripled since 2001, reaching an estimated 7.3 million students studying outside their home countries by 2024 \cite{iom2024internationalstudents}. This significant growth has made international education a global market worth over USD 500 billion a year, with the Asian economies, particularly China, India, Pakistan, Bangladesh, and Vietnam, supplying most of the world’s outbound demand \cite{simonds2023understanding, unesco2026trends, zamanFactorsShapingBangladeshi2023, ahmed2025braindrain, fenghorta2021}. Consistent with this growing demand, recruitment agents and institutional recruitment teams influence are increasing, making the process as much a commercial transaction as an educational one \cite{robinsonpant2018recruitment, beech2018migration, yang2025takenforgranted, salem2025perceptions}. Migration scholars call the systems that carry this movement an infrastructure, making mobility easier for some and harder for others through linked technologies, institutions, and actors that shape whether and how people move \cite{xiang2014}.

In Bangladesh, the number of students outbound for higher study has more than doubled over the decade from 37,332 in 2013 to 76,328 in 2023 \cite{unesco2026outbound}. In most households, migration is a livelihood strategy rather than a lifestyle choice, depending on social and financial capital before a child ever leaves, sometimes discussed nationally as a brain drain concern with remittance effects extending beyond the individual student \cite{rashid2024social, ahmed2025braindrain, islam2025migration}. Thus, many commercial structures exist in the form of education consultancies to help students make this important decision. Alongside this infrastructure, there are volunteer-run Facebook groups that a growing share of aspirants turn to instead. While previous research highlights the aspirants' experiences in these groups, the moderators and admins voluntarily facilitating the service remain unstudied \cite{jayadeva2020keepcalm}.   

Previous HCI scholarship on volunteer moderation largely focused on pseudonymous Western platforms. Studies on moderation practices on Twitch \cite{wohn2019}, Reddit \cite{dosono2019moderation}, Wikipedia \cite{menking2015}, live-streaming teams \cite{caiwohn2022}, Facebook groups \cite{gibson2023} describe it as emotional and civic labor \cite{matias2019civic} that can be redesigned around community-based, self-moderating models \cite{seering2020}. The rising cost can push back moderators to withdraw or retreat into anonymity \cite{dosono2019moderation, wohn2019}. Further works focus on understanding motivation for taking volunteer work, offering accounts grounded in empathy toward those in need \cite{batson1981}, internalized obligation \cite{schwartz1977}, generalized reciprocity \cite{kimReceiversGiversUnderstanding2022}, and the private satisfaction of giving itself \cite{andreoni1990}. These accounts are extended to a functional view of volunteer motivation \cite{clary1998} rooted in socialization and multilevel prosocial behavior \cite{clary1986, penner2005} -- an impulse that sustains without any formal obligation \cite{gibson2023teams}. Other studies show the true cost of this unpaid labor. Star and Strauss \cite{star1999} describe invisible uncredited articulation work that holds systems together without ever counting as “real” work. Recent HCI work on informal exchange systems from the Global South has focused on money, appropriation, and commissioned labor \cite{Mwesigwa2024}, leaving unpaid, uncompensated care work in these systems unexamined.

Taken together, this body of work treats volunteer moderation in educational groups and paid educational guidance as separate problems. High-stakes educational guidance, and the role of paid agents within it, sits outside HCI literature altogether. Nor has HCI examined the platforms where unpaid moderators take up this work themselves: real-name, high-stakes, Global South spaces where moderators do agent-like work, helping people make irreversible decisions about their futures using nothing but information and goodwill. Unlike support groups built around exchanging feelings or offering suggestions, what moderators deal with here is someone's aspiration to lead a different life \cite{appadurai2004}, sustained through what Miyazaki \cite{miyazaki2004b} calls a method of hope: a way of orienting action toward a future that is not yet certain, where getting it wrong can carry lasting consequences for the future that method was meant to secure. Yet HCI has largely looked past this labor. The field's own review of a decade of migration research confirms the gap that it has studied migrants' technology use extensively, but rarely the unpaid infrastructure that sustains their journeys before they even leave \cite{sabie2022}.

We take up this gap through \textit{HigherStudyAbroad: A Global Hub of Bangladeshis}, one of the first groups of its kind built to support Bangladeshi higher-education aspirants, examining in depth the role its volunteer moderators play and how they position themselves within this space. We ask what motivates moderators to stay and to build a sustaining digital infrastructure for aspiring students, not the one-off generosity typically described in prior literature. We ask the following questions:

\begin{itemize}
    \item \textbf{RQ1}: What motivates moderators keep their commitment to voluntarily sustain the group?
    \item \textbf{RQ2}: What visible and invisible labor does the group's operation require?
    \item \textbf{RQ3}: How does the group position itself as a viable free alternative to paid agents and consultancies?
\end{itemize}

To address these questions, we conducted a qualitative study involving semi-structured interviews with 17 volunteer administrators and moderators of the Facebook group, \textit{HigherStudyAbroad: A Global Hub of Bangladeshis}, analyzed using thematic analysis. Our findings show that moderators are driven by altruism: a deeper empathy for others that combines a sustained concern for others' welfare with an intrinsic sense of fulfillment moderators derive from the act of helping itself. This altruism is neither a debt being repaid nor a general act of reciprocity. Many moderators had once relied on the group themselves, but their continued labor reflected a commitment that outlasted any personal stake in the outcome. Over time, this altruism became the organizing logic of the group, sustaining its operational, epistemic, coordination, and emotional-temporal labor and lending it legitimacy against paid consultancies. Most of this work remained invisible and unsupported, sustained not by policy or pay but by moderators' quiet care for strangers they would likely never meet.

We make three contributions to HCI literature through this work. First, we extend altruism and volunteer-moderation scholarship, developed largely on pseudonymous Global North platforms, to a real-name, Global South migration setting where the stakes of participation are high. We show that in this setting, altruism operates as an organizing logic rather than merely a starting motivation. Second, we provide an empirical account of the invisible labor this logic sustains, unpacking it into operational, epistemic, coordination, and emotional-temporal forms. Third, we show how the group draws on altruism to establish legitimacy against paid agents and consultancies. We close with design implications grounded directly in what our participants envisioned.

\section{Study Context}

\textit{The Group.} HigherStudyAbroad: Global Hub of Bangladeshis, known simply as HigherStudyAbroad (HSA), is a private Facebook group founded in 2010 \cite{hsabdfb} with a mission to support Bangladeshi students pursuing higher education abroad, especially at the graduate level. As of August 2026, the group had about 226,900 members and about 30 volunteers. They work in two main roles:

\begin{itemize}
\item \textbf{Admins:} Define community guidelines, train and onboard new moderators and manage moderation in general.
\item \textbf{Moderators:} Supervise day-to-day operations like reviewing posts, reviewing membership requests, and moderating content.
\end{itemize}

Admins and moderators have already been through this process. 
Many have completed their Master’s or PhD degrees and are now overseas. They actively advise, guide and inform prospective students. They also keep the group as a bridge between students coming in and students already abroad for information on applications, funding, admissions and visas. The combination of large-scale peer support, unpaid moderation, and platform-mediated maintenance makes \textit{HigherStudyAbroad} a useful setting to study what sustains voluntary contribution and how such communities position themselves against commercial alternatives. 

\textit{Origins and growth.} The group was started because the founder experienced a gap himself, not based on a plan. He came to the US in August 2010 for his own Master’s degree and found that Bangladeshi students had no knowledge networks. He observed that the other South Asian students from the area were profoundly linked. An applicant would know his or her final placement before leaving home. But Bangladeshis were fragmented and disconnected, with alumni networks spread out. He started this Facebook group in 2010, the same year he migrated to the US, which is believed to be the first of its kind for Bangladeshi students. The group grew to about 500 people through informal word of mouth in only five or six months.

The founder was the only active administrator of the group for most of 2010 and 2011. Over time other admins and moderators organically joined. Many first met the founder at seminars local students personally arranged during his return visits to Bangladesh, where local volunteers who had helped with logistics were later recruited into moderation roles based on their prior engagement. He remains group's founder and de facto administrator to this day.

\section{Related Work}
In this section we review the existing literature on three areas. We start with prosocial motivation and its development. We then discuss invisible labor, moderation tools, and the role of AI in such volunteer-run settings. Finally, we consider the migration infrastructure and commercial market that our study site faces.

\subsection{Prosocial Motivation and the Shift from Receiving to Giving}

Batson \cite{batson1981} demonstrates that ethical concern serves as the sole reason for assisting strangers, even in cases where the personal cost of disengaging is nonexistent. Similar findings were reported in another study, showing that individuals have a strong internal moral obligation, comparable to personal commitment, that drives them to action \cite{schwartz1977}. This is further supported by longitudinal evidence from a study on AIDS volunteers \cite{gibson2023teams} and a previous study by Clary and Miller \cite{clary1986}. The “warm glow” effect described by Andreoni explains the intrinsic satisfaction derived from helping, independent of the outcome generated \cite{andreoni1990}. All of this work together provides a complete story of volunteer motivation, where helping is an ever increasing quality, not a fixed trait to be cultivated and passed on.

This story is further evaluated in online spaces. Ren et al. identified two distinct pathways to attachment. People are loyal to an identity of a group or they are loyal to specific people in it \cite{ren2007,ren2012}. This is what Kollock termed a gift economy. People give first and trust that some return will come, someday, from someone \cite{kollock1999economies}. The pattern we see in our own group is similar to the study done by Kim and Wang on a Korean support group. Members came in needing assistance and they slowly turned into givers once they had felt the hardship enough times to think it was theirs to hand down \cite{kimReceiversGiversUnderstanding2022}. Yao et al. examined a comparable transition in a cancer support community. Even longtime givers found it difficult to keep going and some quietly left after too many losses \cite{yao2021joinstaygo}. In their previous work on the same platform, they found that commitment looks different for a newcomer than for an old-timer, so the same community ends up asking two different things of two different people \cite{yang2017commitment}. These studies trace the trajectory from receiver to gift-giver mostly at the level of an individual’s changing role within a community. In our study we examine a group that has taken this same journey and made it the purpose for the entire organization.

\subsection{Invisible Labor, Moderation Tools, and the Uneven Promise of AI}

Once an individual becomes a helper, their help transforms into legitimate work, with the majority of this work remaining concealed. Star and Strauss called this articulation work \cite{star1999}, the quiet coordination that keeps any group task together. Matias studied Reddit moderators, calling their unpaid effort civic labor, work that keeps public conversation going even when almost no one notices \cite{matias2019civic}. Wohn, and separately Dosono and Semaan interviewed volunteer moderators directly and heard the same thing in different words that the job requires real emotional labor, not just clicking buttons [32, 42]. Reddit has a problem that Facebook Groups does not. Kuo et al. interviewed twenty six Facebook Group moderators and found that the tools available to them had not kept up with the scale of these groups \cite{kuo2023unsung}. Using three different automated tools \cite{kuo2023unsung}, they found that it was not the accuracy of the tool that mattered most, but rather whether the same moderators retained ultimate control over the final decision. Gibson found a similar coping mechanism, that Facebook moderators tend to share the emotional toll of their work across multiple people rather than shouldering it alone \cite{gibson2023teams}. Sch\"opke-Gonzalez et al. showed what happens when that support erodes, that moderators leave primarily because of conflict with their own co-moderators as opposed to abusive users \cite{schopke2022}.

Large language models are starting to enter this same space, and the early results are mixed rather than reassuring. Kumar et al. asked GPT-3.5 to apply real subreddit rules, and found that it works well for some communities but poorly for others, without clear reason why \cite{kumar2024watch}. Kolla et al. took things a step further, and tested an LLM moderator against real moderator decisions. It caught most of the harmless posts but missed most of the actual violations, especially those that needed context \cite{kolla2024llmmod}. Huang argues that the whole comparison asks the wrong question, as content moderation is really a form of governance, and governance survives on legitimacy and explanation, not just correct answers \cite{huang2025legitimacy}. This work shows that moderation labor is heavy, mostly invisible, and only partially mitigated by the tools designed to reduce it. It says less about how a bunch of volunteers, already shaped by a decade of imperfect tools, decides what to trust next, and what to keep doing by hand anyway.

\subsection{Migration Infrastructure and the Market Altruism Stands Against}

The group we study is positioned within a larger system that it never chose, a national migration pipeline and a market of paid education agents. Migration is enabled and conditioned by a web of technology, institutions and people rather than one person just deciding to go \cite{xiang2014}, argue Xiang and Lindquist. Much of that web now runs through social media, allowing people to build support networks before they even leave home \cite{dekker2014}, although Dedecek Gertz found these connections mainly form around vocational training, language certification, and school logistics rather than university access \cite{gertz2023migration}. Robinson-Pant and Magyar describe education agents in the paid sphere as a commercial broker and cultural guide, profiting from a student’s move, but also shaping that student’s views of the destination country \cite{RobinsonPant}. Titmuss compared blood donation in the UK and US, and found that a paid system produced a lower quality and more wasteful supply than a system based on voluntary giving \cite{titmuss1970}. Gouldner's norm of reciprocity, the idea that people help those who have helped them, suggests that unpaid help can come with a different kind of obligation than a paid transaction \cite{gouldner1960}. This literature shows that migration is based on infrastructure, and that commercial infrastructure has a real cost. But it does not go all the way to a group based entirely on the opposite, altruism as the entire operating structure, holding its ground next to the paid agents working the same territory. Our study examines exactly that group.

\section{Method}

We conducted a qualitative interview study with current and former admins and moderators of \textit{HigherStudyAbroad}. We focused on their experiences with volunteering and community maintenance, alongside the influence of technology on their work.

\subsection{Participants Recruitment}

We recruited 17 members of the moderation team. Admins and moderators were based in different countries around the world. Their experience with the group ranged from 5 to 17 years. All participants identified as men. Among those who reported their age, participants were between 30 and 40 years old. \hyperref[tab:participants]{Table~\ref*{tab:participants}} provides an overview of the sample.

\begin{table*}[t]
\centering
\caption{Overview of interview participants. ``N/A'' indicates information not reported.}
\label{tab:participants}
\small
\renewcommand{\arraystretch}{1.15}
\setlength{\tabcolsep}{5pt}

\begin{tabular}{@{}lllllll@{}}
\toprule
\textbf{ID} & \textbf{Role} & \textbf{Gender} & \textbf{Age} & 
\textbf{Field} & \textbf{Current Location} & \textbf{Years with Group} \\
\midrule
P1  & Moderator        & Male & 36  & Science                  & USA        & 7    \\
P2  & Founder, Admin   & Male & 39  & Science, Business        & USA        & 17   \\
P3  & Admin            & Male & 35  & Social Science           & Canada     & 9    \\
P4  & Admin            & Male & 32  & Science                  & Bangladesh & 10   \\
P5  & Admin            & Male & 32  & Science                  & Australia  & 12   \\
P6  & Moderator        & Male & N/A & Science                  & USA        & 8    \\
P7  & Moderator        & Male & N/A & Science                  & USA        & 7    \\
P8  & Moderator        & Male & 31  & Science                  & USA        & 10   \\
P9  & Admin            & Male & 36  & Science                  & Singapore  & 7    \\
P10 & Moderator        & Male & 34  & Science                  & USA        & 9    \\
P11 & Moderator        & Male & 30  & Business                 & USA        & 6.5  \\
P12 & Moderator        & Male & 31  & Science                  & Australia  & 6.5  \\
P13 & Former Moderator & Male & 38  & Science                  & USA        & 6.5  \\
P14 & Moderator        & Male & 40  & Science                  & USA        & 11   \\
P15 & Moderator        & Male & 32  & Business, Social Science & USA        & 6    \\
P16 & Moderator        & Male & 30  & Science                  & USA        & 5    \\
P17 & Moderator        & Male & 32  & Science                  & USA        & 9    \\
\bottomrule
\end{tabular}

\Description{The sample comprised 5 admins (including the group's founder),
11 moderators, and 1 former moderator, spanning founding members to more
recent moderators (5 to 17 years of involvement). Participants were
predominantly from STEM academic backgrounds, with a smaller number from
business and social science fields. Participants ranged from 30 to 40 years
of age, and all participants who disclosed gender identified as male. They
were geographically distributed across the United States, Canada, Bangladesh,
Australia, and Singapore, reflecting the group's transnational diaspora
structure.}

\end{table*}

The authors were general members of the group before beginning this study. When we decided to study the group's admins and moderators, we identified them through the admin and moderator badges displayed on their Facebook profiles and began contacting them directly via Facebook Messenger to request interviews. Four members initially volunteered, including the group's founder. Through them, one moderator introduced the study to the team's private Messenger group, using an invitation prepared by the first author. This yielded five additional participants. We then continued snowball sampling through these participants, while simultaneously continuing to message other moderators directly in their inboxes. Several people we approached could not participate because of academic, professional, or family commitments. Participation was voluntary and uncompensated.

\subsection{Data Collection}

We conducted semi-structured interviews to gather data from the admins and moderators. At first, a pilot interview was conducted with a moderator to refine the interview protocol. There were no substantive changes made to the interview protocol after the pilot interview and this interview  data is included in the final dataset. The authors conducted the interviews between June and August 2026 all through Google Meet. Interviews lasted approximately 50 minutes on average. Four were conducted in English, and the remaining thirteen primarily in Bangla, participants chose whichever language they were most comfortable with. We audio-recorded each interview with participants' consent.

To begin, we asked all participants about their journey of higher studies and involvement with our group. We then talked about why they volunteer and how that motivation developed over time. In the later questions, we talked about the labor of moderation, team coordination, member interactions, tough moderation calls, and burnout. We discussed Facebook moderation tools and other digital tools employed by the team. Lastly, we inquired about the paid education agents and consultancy services offered and the changes in the higher-study information landscape, which includes the increased access to online information and generative AI, along with the participant’s wider reflections on migration. The semi-structured format allowed participants to introduce experiences that were important to them.

Our dataset consisted of the interview recordings, transcripts, translations, and research notes. The researchers took notes during and after interviews to record emerging observations. These notes also helped us identify topics that required further exploration in later interviews. The first author manually transcribed all recordings. She transcribed the Bangla interviews in Bangla before translating relevant material into English. The research team reviewed translated quotations when clarification was needed.

\subsection{Data Analysis}

For data analysis, the first two authors, who are native Bangla speakers, translated and transcribed the audio recordings manually. Analysis began alongside data collection, providing a context-based understanding that informed subsequent interviews. We used thematic analysis~\cite{braun2006using} and an inductive, open-coding approach~\cite{corbin2008basics}. All three authors coded each transcript individually, labeling data segments with their own codes. The labels each author produced were then compared. When labels matched, they were accepted as consistent. When labels did not match, the whole team discussed that piece of data until an agreement was reached. Example codes include `altruism and paying it forward', `invisible moderation labor', `positioning against paid consultancy agencies', 'emotional labor under public scrutiny', and `AI relevance to the group's future'. We did not use predetermined themes, in order to maintain openness to emerging patterns in the data. All three authors observed that, after approximately 8 interviews, subsequent interviews yielded largely similar codes. The whole team then worked together to group related codes into broader themes that best illustrate our findings.

\subsection{Ethical Considerations}

The study received institutional ethical approval from the University. Before every interview, we explained the study's purpose, how the data would be used, and that it was voluntary and could be withdrawn at any time. We took informed consent prior to recording. Throughout the paper, participants are referred to as P1 – P17. Because we name the group itself, some participants, especially the founder and long-serving admins, may still be recognizable to those familiar with the community, even with these identifiers in place. Participants were aware of this and agreed to have the group named and their roles described. We also shared relevant excerpts with them before submission to make sure they were comfortable with how their contributions came across. We did not collect Facebook activity logs or analyze participants' posts, comments, or private messages. Everything we report comes from the interviews themselves.

\section{Findings}
Our findings shows a sophisticated, voluntary-run online infrastructure built to help Bangladeshi students in navigating the intricate landscape of pursuing higher education abroad. We found five interconnected themes, with altruism being the primary driver for running this higher studies group. 

\subsection{From Altruism  to a Platformed Culture}

\subsubsection{Altruism as the Central Motivation}

Altruism is the strongest motivation for our participants as they rarely described materialistic gains such as recognition, professional advancement, or future return as the impetus for their voluntary labor. Instead, helping others genuinely made them happy and instilled a sense of fulfillment. As P1 said, \textit{ ``Whatever I had been through, if others are going through the same situation, I would like to help them.''}. Similarly, P7 characterized helping others as \textit{``totally natural''}, while P1 described their motivation as, \textit{``Altruism itself is a reward---helping others feels good.''} P14 expressed similarly that he simply has a \textit{`volunteer-spirit'} and he is someone who naturally loves to help people.

More notably, digital platforms provide them an opportunity to extend this disposition beyond people they already know. The knowledge acquired through one person's application journey could be documented once, posted publicly, and later accessed by many applicants. P2 described the asymmetry between the efforts required to find information and share those online:

\begin{quote}
\textit{``I dedicated significant time to personally gather all the relevant information. I thought about it later, if I share the answers I already know, it doesn't take much time. But for others, it is a huge help.''} (P2)
\end{quote}

Our participants repeatedly said that they would like to use this knowledge to reduce uncertainty among new applicants. P6 showed concern in his own words, \textit{``didn’t want other people to go through confusion and difficulties''.} They were particularly interested in helping underprivileged applicants who lack information. As P12 said,

\begin{quote}
\textit{``I did not get any help from anyone, I researched by myself using various available online resources or websites. But not all students have the same capacity -- so I always feel that if someone has the aspirations for higher education from an underprivileged background, then they should be given the chance. I always stand by them to help those who are swimming in the ocean in the absence of a direction to find''.} (P12)
\end{quote} 






The founder (P2) recalled that he used to spend four to five hours daily answering questions back in 2010 and at times would wonder why he had been doing this for free. He shared this with his mother:

\begin{quote}
\textit{``She asked me, `How much time do you think you could have saved if you had someone to guide you?' I said, 'Maybe one year.' Then she asked, `How many people are there in the group now?' I said, `Around 500.' Then she said, `See, if you are helping 500 people at present, then you are pushing Bangladesh 500 years ahead.' Since then, every higher study aspirant I know about, I try to give back one year to them. That's pretty much my fundamental motivational factor''.} (P02)
\end{quote}

In his response, P13 offered a more nuanced perspective on altruism:

\begin{quote}
\textit{``Performing any altruistic work is itself selfish ... as it makes us happy. Whenever we do altruistic work, we do it for our own wellbeing. Helping others makes us self-fulfilled, and that is a very selfish reason''.} (P13)
\end{quote}

Some of our participants explained that they had received substantial support from this group as an applicant, which motivated them to become a moderator of the group later. For example, P13 expressed this motivation as \textit{``pay that forward, on a much larger scale.''} P17 echoed this sentiment, joining as a moderator to lend `a helping hand' to as many people as possible. In summary, all of our interviewed moderators found this group as an outlet for selflessly helping new higher education aspirants, without expecting anything in return and by giving up their own time and energy.  

\subsubsection{From Individual Helping to a Platformed Collective Identity}

Over the years, altruism had gradually fostered a shared identity among the group moderators. They call themselves `altruists' and the group took on the motto `harbingers of selfless advising.' P11 explained how this common orientation shaped their relationship with one another:

\begin{quote}
\textit{``Since this is an altruistic thing, individual ego doesn't have a place here. We do not compete with one another... Each one of us is trying to give as much as we can from our end.''} (P11)
\end{quote}

Facebook played a very important role here as the most dominant social media platform in the country. During the early days, the moderators participated in physical seminars at different Bangladeshi universities. Gradually, they got geographically dispersed but still continued their commitment to the cause via this common group. P13 described this 
experience: 

\begin{quote}
\textit{``Earlier we used to go to universities and run these seminars, everywhere. But once we had come here ourselves, we couldn’t do that anymore. But we still wanted to help people, so here is our group.''} (P13)
\end{quote}

This digital altruism network gradually evolved over time to become an inclusive, sustained digital archive. In addition to helping existing aspirants, the moderators also wanted to include students who are not much familiar with the concept of higher study abroad and pique their interest about its potential. P13 explained how they facilitated the onboarding process by creating a repository of basic terminologies before these new group could learn more from other members.

\begin{quote}

\textit{``If one is unfamiliar with the concept of higher study, they can start here by learning the basic things such as what is a bachelor’s program, what is a post graduate degree, how many types of funding are there, Fall vs. Spring semester abroad, what does ‘I-20’ mean, and build upon that according to their own interest.''} (P13)
\end{quote}

The above responses indicate that the potential of Facebook as a ``networked public'' allowed altruistic efforts to transform into persistent, collective, globally distributed, searchable, and scalable efforts by our participants. 

\subsection{Making Altruism Sustainable: Human Labor and Platform Tools}

The admins and moderators combined human labor with Facebook's moderation features, Messenger, digital knowledge resources, and other tools to sustain their altruistic activities. We identified four forms of labor in this regard: (i) \textit{operational labor} that keeps the group functional and safe, (ii) \textit{epistemic labor} that maintains useful and reliable information, (iii) \textit{coordination labor} that organizes distributed volunteer contributions, and (iv) \textit{emotional and temporal labor} through which moderators absorb the personal demands of this work. Table~\ref{tab:labor} summarizes these forms and their associated work details. While technology has minimized certain repetitive duties, it has required the review, interpretation, coordination, and maintenance of similar tasks. 
\begin{table*}[h]
\centering
\small
\renewcommand{\arraystretch}{1.4}
\caption{How human labor and digital tools jointly sustain \textit{HigherStudyAbroad}.}
\label{tab:labor}
\begin{tabular}{@{}p{2.1cm} p{4.2cm} p{3.6cm} p{1.4cm}@{}}
\toprule
\textbf{Form of Labor} & \textbf{Human Work} & \textbf{Digital Tools} & \textbf{Visibility} \\
\midrule
Operational & Approving members and posts, removing fake accounts and spam, updating group files. & Facebook approval controls, Admin Assist, keyword triggers & Visible \\
\addlinespace
Epistemic & Interpreting posts, judging credibility, identifying relevant expertise. & Group search, document archives, pinned answers & Invisible \\
\addlinespace
Coordination & Dividing tasks, training new moderators, resolving ambiguous cases. & Messenger group, Google Meet, polls & Invisible \\
\addlinespace
Emotional and temporal & Managing conflict, workload, and burnout across an unpredictable schedule. & Persistent notifications, shared inboxes & Invisible \\
\bottomrule
\end{tabular}
\end{table*}

\textbf{\textit{Operational labor}} is the most visible layer. The moderators mostly review membership requests and posts, answer questions, remove inappropriate contents, and manage promotional activities. Considering the scale of the group, even routine platform actions became demanding. P7 described the membership approval procedure:

\begin{quote}
\centering
\textit{``Sometimes the number gets quite high, crossing a thousand... looking at a member's ID and then analyzing it to figure it out -- that usually takes at least 60 to 90 seconds ... member approval feels a bit overwhelming to me.''} (P7)
\end{quote}

Earlier moderators had even created their own technical solutions when automation was not available. P9 recalled:

\begin{quote}
\centering
\textit{``During that time, we didn't have much access to automation tools, so we did it manually. And I would sometimes, personally, automate things a bit myself. I would write, for example, a PHP script, and approve ten or fifteen at once that way... But not everyone was technical enough to do that, they mostly did it manually.''} (P9)
\end{quote}

With the passage of time, new platform features gradually made these tasks much easier. P13 described using Facebook's built-in \textit{Admin Assist} feature:

\begin{quote}
\centering
\textit{``There are a few Facebook built-in tools we use nowadays. For example, there's an Admin Assist feature with some tools, and we use those to filter who can request to join the group ... We can also decline certain posts if they don't meet certain criteria... Certain words are trigger words we've banned as well, we just want to keep the group clean.''} (P13)
\end{quote}

P7 similarly described configuring Facebook to alert moderators when particular words appeared in posts or comments. These tools enabled moderators to encode community rules into platform features. Nevertheless, participants still wanted much stronger technical solutions:

\begin{quote}

\textit{``If a feature exists for detecting bot accounts on Facebook with a built-in feature to automatically reject those, that would be great. For example, if there are 300 bot accounts and we can reject them all in one shot, that would improve things.''} (P7)
\end{quote}

At the same time, moderators resisted thinking about automation as a replacement for human contemplation. Human judgment is another moderation tool, as P13 described:

\begin{quote}
\centering
\textit{``Another tool that we use is our brain. When you are looking at a thousand profiles every day, you immediately know which one is a fake profile and which one is real. We use our brain like that.''} (P13)
\end{quote}

The moderation system is therefore hybrid at its current stage. Although Facebook could filter some accounts ahead of time, expose trigger words, facilitate approval or removal, the human labor still plays a significant role through manual reviews and interpretations. \textit{\textbf{Epistemic labor }}is a significant layer of this interpretive work. P6 separates what members saw from the judgments behind it:

\begin{quote}

\textit{``The visible work includes reviewing membership requests, checking whether their profile and answers appear to be genuine, approving relevant posts, removing spam and promotions, etc. But there is much more invisible work. For example, read the entire thread and understand the context before approving or removing anything. This is the most difficult part.''} (P6)
\end{quote}

Several participants describe similar situations. P3 described the same weight of the work, recalling how the demands of the role became clear only once they took it on themselves. Approving a single post, they explained, requires reconstructing the full context of what a member is actually asking or offering before any decision can be made.

\begin{quote}
\textit{``Approving posts might sound like small work, but it's not. When I started doing this myself, that's when I understood how much effort it actually takes. I have to read each post and understand what's really being asked or shared, especially with visa interview experiences, member profiles, and genuine scholarship suggestions. A few enthusiastic members also share immigration policy updates. It's not simple. It takes a lot of time. Once I started doing this, I realized it's actually tremendous work.''}
\end{quote}

This weight of interpretive work led some moderators to imagine automated support rather than accept it as permanent. P15 explicitly suggested a simple summarization tool for exactly this bottleneck. He said, \textit{``Maybe a AI that give us a short summary might be a help.''} (P15)
The suggestion is narrow by design. P15 is not asking AI to decide what gets approved, only to shorten the reading a moderator must do before making that decision. This matches what P3 and P6 both described as the hardest part of the work, understanding a post's full context, and points toward the design direction we return to in our Discussion. Besides, our participants reported that they have to constantly switch between group posts and inbox messages, as group members often struggle to select the appropriate method of communication. Despite having all the resources in the group itself, members often make requests through inbox messages and repeatedly ask them to check the inbox by commenting on different posts. However, admins and moderators clearly prefer public responses over private inbox messages as those remain searchable and useful to future applicants. As P1 described:

\begin{quote}

\textit{``The requests in the messenger inbox were quite overwhelming. If someone messaged us, we tried to redirect them to the group resources. If someone in the group commented `please check inbox, I have a query', we would make them post the query publicly, so that others could also get the answer by seeing our response.''} (P1)
\end{quote}

Our participants also reported that they divide their workloads based on individual expertise and experience of the subject matter. For instance, P7 explained that he redirected Canada-related questions to an admin living there because \textit{``I actually don't have much to offer here.''} The admin would later redirect a USA-related question to another member, and so on. This effort relies on \textbf{\textit{coordination labor}} that mostly take place outside the visibility of regular members. P9 described an infrastructure of Messenger groups, smaller volunteer groups, internal knowledge bases, calls, and Google Meet sessions among the admins and moderators. During busy periods, volunteers form specialized teams:

\begin{quote}
\centering
\textit{``During peak time what we do is dividing people into sub-groups. Some update knowledge articles, some handle the comments, we also had maintain a repository where we keep answers to common FAQ-type questions.''} (P9)
\end{quote}

Time zone coordination is also technologically mediated for running the group seamlessly. Our participants described using polls to identify meeting times for moderators distributed across Bangladesh, Europe, Australia, Canada, and multiple US time zones. They view this geographic distribution as operationally useful because moderators could collectively process requests during different hours of the day to ensure continuous availability. 

Eventually, the continuity of digital access result in \textit{emotional and temporal labor}. Moderators are instructed to check every queue, inbox message, reports, and requests upon logging in. P9 said:

\begin{quote}

\textit{``there were weeks when I spent five or six hours a day, and many times I compromised my personal works to do it. I really did that. And in Dhaka, back then, I'd go from my home to a particular literature institution just to fix an event venue -- I'd lose two and a half to three hours just in traffic.''} (P9)

\end{quote}

Managing a group of 226,900 members can bring temporal burnout and create extra work managing disagreements and conflicts. Even from the initial days the admins and moderators had to manage such situations and here comes the \textbf{\textit{emotional and temporal}} labor. P2 described this work as:

\begin{quote}

\textit{``We had to be very hands-on in the early days. There was no approval system back then, posts would just come in, and if someone was being mean or aggressive, they would get a warning. There was a lot of drama back then, even now, though it has reduced since we can filter now. Ultimately we manage these, and that's another work.''} (P2)

\end{quote}

P7 explained that members can `report to admin' when someone is misbehaving, such as posting negative or aggressive comments, after which an admin reviews and takes action. But identifying the issue is just the beginning. Handling such situations can cause burnout. P12 described feeling it himself and watching it happen with others, for some admins had once been highly active, giving 6 to 7 hours a day to the group.

Participants also described friction with platform design itself. P7 said Facebook offers no way to flag or archive outdated posts, so moderators periodically review threads and remove old ones manually, since stale visa requirements or scholarship deadlines left visible can be mistaken for current information. A related friction came from Facebook's comment ordering. Members wanting to follow a post would comment just the letter ``f,'' a common Facebook-group convention, but this bumps the post back to the top of the feed as its most recent activity.

\begin{quote}
\centering
\textit{``A lot of people write `f' in the comments when they want to follow a post, but that's not ideal, because every time someone does that, the post gets bumped back to the top since it becomes the most recent comment. We've repeatedly asked people to use Facebook's built-in `Follow' option instead. Maybe a tool could help, where whenever someone comments `f,' it gets automatically deleted and the person just follows the post instead. The `f' comment doesn't contain any answer, and when you have hundreds of them, it becomes like looking for a needle in a haystack.''} (P14)
\end{quote}
Participants traced both frictions to the same cause. Facebook favors keeping people engaged over surfacing relevant content, so a post kept alive by a stream of ``f'' comments can rise above an older post that actually holds the answer someone needs.

In summary, repetitive tasks are minimized by automated filters, searchable archives make answers reusable, information digitization speeds up distributed coordination, and reporting tools help surface conflicts for review. However, these affordances still rely upon human labor spent in interpreting contexts, managing exceptions, maintaining resources, resolving conflict, coordinating between different time zones, and absorbing the burnout that comes from sustaining all of this over time.

\subsection{Building a Credible Non-Commercial Information Infrastructure}

\subsubsection{Competing Through Free, Experience-Based Knowledge}

Participants find \textit{ HigherStudyAbroad} positioned within a digital information ecosystem that incorporates university sites, social media, higher education agents and consultants, and increasing usage of AI. However, they distinguish their service from others because their guidance is free, public, crowdsourced from many contributors and more accurate, and based on firsthand experience. P11 feels their credibility stems from the fact that their service is free:

\begin{quote}
\centering
\textit{``There’s no bargain or deal in this context, or anything else for that matter. Everyone can benefit, but no one here is charged.''} (P11)
\end{quote}

Several participants noted that working through an education consultancy or agent added friction to the application process, largely due to the formality of communication whereas the group has a very informal structure and a friendly environment. P3 explained that when a consultancy handles the application, students are not aware of every step of the process, resulting in a lack of transparency since a third party is completing the work on their behalf. Alongside there is monetary transaction so the entire process becomes business. P7 compared this to commercial incentives:

\begin{quote}
\centering
\textit{``The commerical agencies have come here to do business right? They might intentionally complicate or overhype certain things because if the process doesn't seem difficult, they can’t justify charging you the amount.''} (P7)
\end{quote}

Unlike one-to-one commercial consultancy, the platform manifest a form of distributed credibility and encourage open participation. An aspirant does not have to depend on a single answer as they can scroll the group, search through different responses, and compare those to make an informed decision. 

\subsubsection{Governing Trust Through Platform Rules and Human Judgment}

Open participation also created risks. Fake profiles, misinformation, commercial promotion, scams, and harmful comments could enter through the same platform that enable peer support. Moderators therefore used Facebook's governance mechanisms to protect the credibility of the community. Various options such as post-approval, deletion, pinning, and account management features became tools through which moderators enacted this principle. P13 described this as information control and an early principle they all followed:

\begin{quote}
\centering
\textit{``Information control was very important from the very beginning. I knew this right from the start, that we don't share any information we don't actually know. We read every submitted post. So if a post contains misinformation, we simply don't approve it. If we see that an answer is good, we sometimes pin it, to show that this is genuinely reliable information. If we see information that isn't good, we delete it.''} (P13)
\end{quote}

However, the platform could not determine credibility by itself. Moderators recognized that higher-study information was contextual. P05 explained that he provided advice only \textit{``to the point of my knowledge at that time''} and tried to gather more context when necessary. Public discussion also allowed other members to correct or supplement information. Commercial activity made the relationship between platform rules and human judgment particularly visible. Moderators generally prohibited paid promotion because they could not verify every provider. P10 explained:

\begin{quote}
\centering
\textit{``If someone says, for example, `I'll take on a student,' then that's fine. But if you say, okay, pay me something. we do not allow that because it's a voluntary group, because we can't actually verify everything. We don't know what's a scam, who's authentic. through our group, we don't want any student to face any scam.''} (P10)
\end{quote}

Yet moderators did not apply this boundary mechanically. P10 recalled one time a general member of a group made some IELTS-resource contents with his personal notes and videos, post that initially resembled promotion. Rather than removing the post outright, moderators brought the case to their private Messenger group for collective discussion, ultimately approving it on the grounds that it could benefit other members. P10 described this decision as:

\begin{quote}
\centering
\textit{``Several of us admins together decided whether it was actually important. We saw he had no intention of promotion. That guy had made some IELTS resources himself, mainly hand-notes and a few video contents, and he was giving away some free materials, not taking any money. So then we thought, okay, it's fine. It's fine to share which will help other students.''} (P10)
\end{quote}

Consequently, the moderation decision shifted across technological spaces. Facebook promoted the posting and offered mechanisms to approve or removal it. Messenger supported collective interpretation where admins and moderators judged the ultimate contributor's intent. P13 also explained why the group used a broad rule against agency posts rather than trying to certify individual businesses:

\begin{quote}
\centering
\textit{``We cannot identify which agency is fraudulent and which is real. We just cannot, we do not have the bandwidth, we do not have the time, the energy, or the effort for that. So, we generally just block all agency-related posts''} (P13)
\end{quote}

Therefore, trust originated from a hybrid governance system. Facebook offers a range of enforcement mechanisms including filters, approvals, deletion, and blocking. The moderators supplied the contextual judgement needed to decide how to deploy those mechanisms. The group's credibility as a substitute stems not only from the free availability of information but also from the ongoing management of the digital realm by volunteers in which this information circulates.

\subsection{Platform-Mediated Mobility and Its Moral Ambiguity}

Although participants mainly perceived of this group as a community that offered educational support, the infrastructures they digitized also contribute to making the pathway of international mobility easier. Accessible data on universities, funding, visas and previous applicants turned what was previously very personal into information available to many more people. Because of this, many participants recognized that facilitating student mobility can also facilitate long-term migration from Bangladesh. P1 argued that international education itself had value beyond migration outcomes. He believed that students could benefit from experiencing different academic and cultural environments:

\begin{quote}
\centering
\textit{``One criticism might be that we are encouraging brain drain. I believe every student should opt for studying abroad at least once, even if they do not settle there. The experience of studying abroad including the culture of different countries, university environment, and facilities.''} (P1)
\end{quote}

This framing separated the community’s immediate purpose from the location of its members. The group’s role was to make opportunities more accessible. In general, participants attributed the decision to remain abroad or return to Bangladesh to the individual, rather than the community that had helped them in achieving that opportunity. This difference was summed up by P5:

\begin{quote}
\centering
\textit{``I see it this way, as a higher study opportunity. Whether they come back to the country later or not, that's a personal matter for the person.''} (P5)
\end{quote}

P6 similarly resisted describing the group's work as simply encouraging people to leave Bangladesh. Instead, he framed access to information as a way of expanding the choices available to students:

\begin{quote}
\centering
\textit{``I would not describe our work as simply encouraging people to leave Bangladesh. We are opening up the world for you. We are giving you a book, and it's your choice whether you burn its pages for fire or use it to harvest knowledge. Our role is to help students access accurate information so they can make informed educational and career decisions.''} (P6)
\end{quote}

This distinction is important for understanding the role participants assigned to the platform. Moderators were able to make information and experience available to a large number of students through Facebook but did not see themselves as determining what students should ultimately do with that information. In the same stance, P15 shared their role in the same way of opening up possibilities rather than prescribing destinations. In this sense, the community served as an information intermediary for mobility and moderators maintained a moral boundary between \textit{enabling a choice} and \textit{making that choice for someone}.

Not all participants had always reflected on this distinction. P9 explained that, particularly during the group's earlier years, moderators were more concerned with helping others achieve the same aspirations they held themselves. The migration consequences of that support remained largely outside their immediate attention. He emphasized once again that the group knew what its role was mainly through aspiration and not through migration: \textit{``Nobody ever thought about this migration aspect from the country. We just saw it as helping people to get their dream.''} His retrospective narrative shows how the meaning of the same volunteer practice could change over time. Offering answers to questions about applications or funding might appear to be a small act of peer support. However, in scale, aggregated posts, resources, and experiences may reduce informational barriers for many pursuing international mobility.

Rather than understanding outward mobility only in terms of brain drain, P2 situated it within the framework of wider economic and demographic constraints. He also linked outward mobility with human resources in Bangladesh:

\begin{quote}
\centering
\textit{``Human resources are Bangladesh's only real economy. What the land itself produces is barely enough for people to survive on. So all Bangladesh can really do is send its people out to serve that need globally.''} (P2)
\end{quote}

 His account complicates a simple opposition between helping an individual leave and helping the country. From this perspective, enabling people to develop their skills and participate internationally could itself carry collective value.

P13 offered another way of resolving this tension. He questioned the assumption that contribution to Bangladesh requires physical presence within the country. Reflecting on his own experience abroad, he argued that digital connectivity and transnational networks allow people to remain involved from elsewhere:

\begin{quote}
\centering
\textit{``It wasn't possible for me to do a PhD in Environmental Science in Bangladesh. In this age of information, social networking, and the internet, you can contribute to your country no matter where you are. I don't think it's necessary to be physically in Bangladesh in order to work for Bangladesh.''} (P13)
\end{quote}

Here, the same technologies that supported outward mobility also made continued transnational contribution imaginable. Participants themselves illustrated this possibility because the moderators themselves helped the Bangladeshi higher-study community while living abroad, using Facebook and Messenger to circulate knowledge back to students in Bangladesh.

Participants therefore did not reach a shared moral resolution about migration. What their accounts make visible, however, is that \textit{HigherStudyAbroad} does more than communicate information. The platform facilitates educational mobility by making experiential knowledge searchable and pairing applicants with a geographically dispersed population. The infrastructure that can facilitate altruistic support can therefore also shape migration pathways at a wider scale.

\subsection{From Information Scarcity to Generative AI: A Changing Division of Labor}

According to participants, the nature of the technological environment surrounding higher-study support has changed substantially since the group's creation in 2010.  At first, Facebook reduced this fragmentation by collecting questions, documents, past experiences, and knowledgeable people, all in one searchable space. This built-up became a big digital archive eventually.

P11 described this transition:

\begin{quote}
\centering
\textit{``Back then, internet availability was comparatively low. So we conducted frequent seminars at different universities when we came to Bangladesh for vacation. But now there isn't much need for that, because information is much more available. Our group is very rich now, because a lot of information has accumulated here over the past 15 to 16 years. All of that has been stored, so now you can access any information you want.''} (P11)
\end{quote}

Generative AI now represents another shift in this information environment. P12 and P16 shared the same thought and noticed changes in group activity. Over the past year and a half, they have seen that the workload, or the number of questions in the group, has gone down quite a bit, because people can now find a lot of information easily through AI. P12 observed that applicants increasingly used generative AI for generic questions and turned to the group for more specific or experiential information.

\begin{quote}
\centering
\textit{``Now people can just ask ChatGPT when they want information. They mostly use the group for insider knowledge. But AI has taken over most of the basic questions, which is good. What's left is people sharing lived experience, the kind of thing you can't get from Google or AI.''} (P12)
\end{quote}

Rather than treating this reduction in demand as a threat, several moderators welcomed it. P16 explained:

\begin{quote}
\centering
\textit{``No one is sure what AI can do now, or what it will be able to do in the coming years. If individuals can find the answers without the group, I don't have any issue with it. If the information is obtained using artificial intelligence, that's fine.''} (P16)
\end{quote}

His response was closely tied to the group's non-commercial structure.P13 expressed a similar sentiment, noting that moderation is unpaid and not a formal job, but rather something they do out of personal interest and no one would object if their workload were to decrease. This response again brings light to the altruistic rationale of the community. Moderators do not need to maintain demand for their service, which suggests a technology that achieved so would still be in the interest of the moderators.

Participants nonetheless acknowledged the limitations of AI. They distinguished generic guidance from recent, situated experience. P10 described this through the US visa application of his younger brother. While ChatGPT was able to explain the DS-160 form, his brother was confused. P10 asked him to look into the document archive and past posts of the group.:

\begin{quote}
\centering
\textit{`I told him to navigate to the group, locate the DS-160 form in the document section, or simply search for DS-160 within the group, since there would surely be many samples available. He searched and found the experiences of different people. Then his confusion was clear. So I would say it's very hard for AI to replace this.''} (P10)
\end{quote}

P10 also added that there was more to it than information accuracy. Having access to the documents and experiences from individuals who had just finished the same procedure instilled confidence that a generic AI response could not. People not only found answers in the searchable social history of the group but also connections. P10 remembers that when he searched the group for ``Idaho State,'' he came across an earlier Bangladeshi student, visited his profile and contacted him directly. He inquired, \textit{``How would AI accomplish this?''} Thus, the platform linked informational search with social discovery.

According to P6, group’s future value will not merely be about supplying information:

\begin{quote}
\centering
\textit{``Today information is widely available through university websites, YouTube, social media and AI tools. Therefore, the value of the group in the future will not lie in providing more information. It will be in organizing, verifying, contextualizing, and personalizing the information.''} (P6)
\end{quote}

Participants therefore described a changing division of labor between people and technologies. Facebook first made peer knowledge persistent, searchable, and scalable, while also creating new moderation work. Facebook's automation tools later reduced parts of that operational burden but continued to require human interpretation. Generative AI and different websites now handle an increasing share of routine information seeking. What participants expected to remain valuable were the tasks that required \textit{recent experience, contextual interpretation, verification, reassurance, and human connection}.

We therefore do not find a simple trajectory in which technology replaces volunteer support. Instead, each technological change reorganizes what volunteers need to do. Altruism remains the motivation, but its technological expression changes, from in-person seminars, to Facebook posts and searchable archives, to platform-assisted moderation, and now toward a possible hybrid relationship with generative AI. The continuing role of \textit{HigherStudyAbroad} lies less in owning information than in helping people interpret, verify, and socially situate it.

\section{Discussion}
Drawing on interviews with 17 admins and moderators of \textit{HigherStudyAbroad}, we examine why they volunteer, how this sustains a credible alternative to paid migration services, and how they see their role in others' migration decisions. We also report how moderators view the future of the group as AI takes on more of the work of distributing higher-study information.

\subsection{Altruism as Motivation and as Infrastructure}
Altruism sits at the center of \textit{HigherStudyAbroad}. Helping aspiring students was not a by-product of running the group but the reason for joining and staying. This is aligned with two well-established accounts of altruistic behavior. First, the empathy-altruism hypothesis posits that people help out of genuine concern rather than reward \cite{batson1981}. Second, norm-activation theory proposes that people help because they feel a personal obligation to do so \cite{schwartz1977}. Both view altruism as a meaningful motivation, not disguised self-interest \cite{piliavin1990}, and a motivation that is more important online, where help can be given without the immediate social feedback of face-to-face giving \cite{lou2023}.

Our findings build on these theories by showing how an individual motivation becomes the operating logic of a community. Admins and moderators described helping as part of who they are, calling themselves altruists and repeating a shared motto that shaped recruitment and how the group presented itself to newcomers. Where the empathy-altruism hypothesis \cite{batson1981} and norm-activation theory \cite{schwartz1977} explain individual motivation, our findings show how it becomes shared, reinforced, and sustained collectively. This extends the warm-glow account of giving, which locates the satisfaction of helping in the individual \cite{andreoni1990}. Here, that satisfaction became part of a shared group identity, echoing the broader observation that online cooperation can persist at scale without the social structures usually thought necessary to sustain it \cite{kollock1999economies}.

The common identity and bond theory explains attachment to online communities as being due to either a common identity shared by the group or bonds with particular members \cite{ren2007,ren2012}. In \textit{HigherStudyAbroad}, however, the attachment to the purpose of the group became more important as admins and moderators migrated across countries and personal ties became more difficult to maintain, mirroring the arc of the Reader-to-Leader framework from peripheral participation to sustained leadership \cite{preece2009reader}. Our findings extend these accounts beyond the issue of why individuals stay. The same altruistic identity provides continuity when the people running the community change and the identity becomes a part of the infrastructure of the group.

\subsection{The Weight of Labor Beyond a Typical Support Group}

In prior HCI literature members of online support groups are often shown sharing emotional and informational support \cite{rubenstein2015, kimReceiversGiversUnderstanding2022}. \textit{HigherStudyAbroad} offers similar support, but with four different types of moderator labor: operational, epistemic, coordination and a combined emotional-temporal labor that encompasses both the affective and time-bound sides of the work. Here responsibility is spread out differently than in a typical support community. The underlying support is created by the members themselves who give advice and experience for free, while moderators and admins work to protect the group from unreliable claims circulating in it. This is why epistemic labor is so important. Moderators decide what information is trustworthy, whose expertise is important and when to step in as members act on those judgments in choosing universities, scholarships and visas. This ensures that a peer-produced knowledge base remains reliable enough for people to build their futures on it.

That labor is also unevenly visible, as Star and Strauss explain in their account of invisible articulation work \cite{star1999, strauss1988}. The same pattern can be seen in Wikipedia’s barnstar system. The peer recognition uncovers a much larger scope of valued work than presented on the platform itself [35]. Facebook has no equivalent. Its tools only indicate a results, a post approved or a comment deleted, never the judgment behind them. Representations of work are political in this way, determining whose effort counts and whose does not \cite{suchman1995}. In HCI literature, even paid crowd workers, whose labor should be easiest to notice and pay for, lose income to unrecognized coordination tasks \cite{toxtli2021}. We observe the same problem for volunteer moderation in a large-scale study of Reddit moderators’ private logs, which found that measuring only visible actions badly understates the labor involved [31]. The admins and  moderators of the group perform this coordination work for free and without any kind of surveillance. Similarly, HCI research on open-source software has found that this invisible coordination grows rather than shrinks as projects scale \cite{geiger2021}, but that work looked at code. But this is an information community  and the invisible labor is trust and judgment of someone else 's future.

That difference produces a different cost. Emotional labor research on moderation usually traces its cost to outside hostility or conflict \cite{dosono2019moderation, menking2015, wohn2019, cai2022coordination, gibson2023}. Recent survey work points to a related but distinct strain, conflict between moderators themselves, harmful behavior from team leads, and simply not enough time \cite{schopke2022}. Our moderators described a cost closer to this second kind. They did not describe the work as personally dangerous or as carrying a stake that could ruin someone's future rather what they named repeatedly was the difficulty of maintaining the credibility of information circulating in the group, since a wrong information can deceive a student, or hinder their progress. Thus, this cost resembles the epistemic labor of verifying facts and the ongoing work of managing disagreement seen in other moderation settings \cite{ren2007, ren2012, star1999, geiger2021}. A recent case study of Facebook Group moderators found the same gap between the scale of the work and the tools available to do it \cite{kuo2023unsung}. What is distinct here is that this labor is carried by a shared altruistic identity, which is what makes moderators willing to keep checking and mediating at real personal cost, even though no tool will ever show anyone else that they did it.

\subsection{Legitimacy Without Payment or Blame}

Moderators build the group's legitimacy on two fronts, against paid agents and against the migration critiques often aimed at platforms like theirs.

Against paid agents, the group's unpaid status is the source of its credibility. Voluntary giving carries social meaning partly because it sits outside market exchange \cite{titmuss1970}, and introducing payment can weaken the motivation to contribute \cite{frey1997}. Being free is more than an economic fact here. Moderators use their lack of financial interest to distinguish the group from paid consultancies, since they earn nothing from recommending a university, destination, or service, and present their advice as guided by the applicant's interest rather than a fee. This performance of disinterestedness also functions as impression management. Research on high-context societies shows that maintaining a credible public image, or ``saving face,'' is itself active labor conducted through ICT \cite{semaan2017saving}. In a South Asian, collectivist setting, admins and moderators are not only proving they lack a financial motive but continuously managing how the group is seen by the community whose trust it depends on. Altruism therefore does three things at once, motivating the help, grounding its trust over commercial alternatives, and standing in as a public reputation the group cannot afford to lose.

This non-commercial identity still operates inside a commercial platform. Admins and moderators are unpaid, while their activity generates engagement and value for Facebook, echoing earlier accounts of free digital labor \cite{terranova2000, fuchs2014} and more recent work on how freely contributed activity becomes captured as economic value \cite{fourcade2020}. \textit{HigherStudyAbroad} is a non-commercial alternative built on commercial infrastructure. Its lack of payment gives it legitimacy with users even as its unpaid labor quietly generates value for the platform hosting it.

The second critique concerns migration itself. Migration researchers often treat the platform mediating movement as partly responsible for its outcomes \cite{xiang2014}, a concern that grows once social media is shown to reshape migrant networks rather than simply carry information across them \cite{dekker2014}. This work also shows migrants building and reworking their identities online as they move, using digital spaces to negotiate who they are across borders rather than just transmitting information through them \cite{lingel2014city}. Our moderators reject the stronger version of this framing. They describe themselves as opening access to information and opportunity, not engineering an outcome, and treat whether someone leaves Bangladesh as the applicant's decision alone, even as the group becomes part of the identity work migrants do before they leave.

This is not defensiveness. Several participants said every student should study abroad at least once regardless of whether they settle abroad permanently, treating the experience as valuable in itself. This view holds up against the brain-drain criticism it responds to. A recent large-scale review finds that skilled emigration from lower-income countries usually raises human capital at home rather than draining it, once remittances, return migration, and knowledge transfer are counted \cite{batista2025}. Earlier modeling work found a related pattern, that whether a country nets a gain or a loss depends on its starting level of education and emigration rate, with the world's largest developing countries, accounting for most of the population affected, tending to net a gain \cite{beine2008}. None of this erases the emotional weight a family may feel when a child leaves, but it means our participants' refusal to see themselves as responsible for migration outcomes is not a rationalization. The decision belongs to the individual. The group's role is to build the capacity to make it, not to make it for anyone.

\subsection{Hope, Human Experience, and the Limits of AI}

Given the labor this work demands, one might reasonably ask why the altruism does not run out. We argue that hope sustains it. Miyazaki's method of hope describes an orientation toward the future \cite{miyazaki2004b}, and our admins and moderators actively build that orientation in applicants who would otherwise lack the resources to build it themselves. A growing thread of HCI and ICT4D work treats hope as something to study directly rather than a background feeling. Recent CSCW work re-reads the method of hope for HCI projects built around moments of collective reopening \cite{ratto2023}, ICT4D scholarship treats information itself as a carrier of hope in under-resourced settings \cite{sahay2023}, and design work with vulnerable populations shows aspirations shaped by what technology makes visible as possible \cite{kumar2019aspirations, dell2016}. Closest to our setting, recent work on internally displaced communities in Dhaka frames hope as an ongoing way of surviving, built through identity and the freedom to leave \cite{safir2025}.

Our case extends this work beyond displacement into everyday, unpaid coordination. Functional theories of volunteering support this too. A single volunteer act can serve several purposes at once \cite{clary1998}, and sustained altruism depends more on ongoing situational conditions than on a volunteer's original disposition \cite{clary1986, gibson2023teams}.

Generative AI is a direct test of this argument. If the altruism here were really about status or being needed, AI should threaten these moderators, since it answers many of the same questions faster and at scale. Instead, our participants converged on one view. AI can absorb basic, repeatable questions, but it cannot replace lived experience. Someone who went through the visa process themselves, made the same mistake being asked about, and worked through a difficult scholarship application with their own strategy carries a weight no model can produce. The group does not become obsolete in their account. It becomes smaller and more focused, centered on the judgment a model cannot manufacture, even as the volume of basic questions it handles drops sharply.

This matches prior work showing volunteer labor adapts to new technology rather than resisting it \cite{geiger2021}, with a direct precedent on Reddit, where algorithmic moderation tools did not replace volunteer moderators but shifted their effort from rule enforcement toward the harder, judgment-based work of building community \cite{he2024}. A separate study across Twitch, Reddit, and Facebook moderators reached a compatible conclusion even before generative AI entered the picture, that active human engagement, not automation, is what actually develops a community over time \cite{seering2019moderator}. This also follows from the non-transactional legitimacy discussed above \cite{titmuss1970}, since a community that earns nothing has nothing to protect from automation, and from the belief in open access that drives unpaid contributors regardless of whether a machine can do the same job \cite{nov2007, bryant2005}.

What stays valuable once AI absorbs routine questions is exactly the layer our typology calls invisible. Interpretation, verification, and reassurance were never simple information retrieval, which matches research on Twitch and Reddit moderation, where a moderator's real value lies in judgment under uncertainty, not fact delivery \cite{seering2020, cai2022coordination, matias2019civic, gibson2023}. Our participants said this more plainly than any theory does. The group will stay relevant. It is only the basic questions that might not.

\subsection{Design Implications}
Our findings surface four specific gaps between Facebook's current tools and what admins and moderators say they need. These are not general recommendations. They come directly from what our participants asked for.

\textbf{Bulk bot detection:} Admins and moderators wanted to flag and decline many fake accounts at once, instead of checking each profile by hand. This matches what moderators on other platforms have asked for, a smarter detection tool, not the simple rule-based filters most platforms offer today \cite{jhaver2019automod, chandrasekharan2019crossmod}. The risk is precision. A bot classifier could wrongly flag real applicants whose accounts are simply new. A safer design keeps human review for borderline cases and automates only accounts flagged with high confidence, so admins and moderators still make the final call.

\textbf{AI summaries of long, detailed posts:} Participants pointed to specific post types that take the most time to review: visa experience threads, profile-sharing posts asking for university or scholarship suggestions, and posts about a particular country's study or immigration policy. These posts are often long and detailed, and moderators must read through the whole post to understand the applicant's background and situation before they can approve it or respond. Prior work on automatic summarization of long user-generated posts has shown that a short, AI-generated summary can let a reviewer grasp the key details without reading the full post first \cite{sotudeh2022mentsum}. A similar tool here, one that generates a short summary of a post's key details (profile background, target country, specific question) as soon as it is submitted, would let admins and moderators quickly grasp the context of each post instead of reading every long post in full.

\textbf{Comment ranking that favors substance:} Participants described a subtler problem short, low-effort comments (simple acknowledgments or "follow" replies) can bury the one comment that actually answers the question. This is a known issue in comment design more broadly, where ranking by engagement can hide the most useful contributions \cite{diakopoulos2011quality}. Ranking comments by informational content, not just reaction count, would keep real answers visible.

\textbf{Flagging outdated information in old posts:} Some participants noted that posts are never archived, so an old policy-related post stays visible alongside current ones with no signal that visa rules, scholarship criteria, or country policies may have since changed. The burden of checking currency falls on the reader. This mirrors a known problem in other knowledge-sharing communities, where old answers often go outdated without any flag, misleading readers who cannot tell current information from stale information \cite{zhang2021obsolete}. A lightweight fix, a visible `last verified' date or an automatic flag on older posts in policy-sensitive categories, would shift this burden from the applicant back onto the platform.

In each case, admins and moderators were not asking the platform to replace their judgment. They wanted it to take over the repetitive parts of the work, so judgment could go where it is actually needed, the same distinction we draw from our findings.

\section{Limitations}
Our study has two main limitations. First, we studied a single group, \textit{HigherStudyAbroad}, chosen for being the first of its kind with the most established moderation practices. Our findings therefore describe one case rather than a cross-section of Bangladeshi study-abroad communities; several similar groups exist, some organized around specific destination countries or undergraduate applicants, and whether altruism functions as an organizing logic there too remains an open question. Our claims should be read as grounded in this specific, high-visibility case rather than generalized across such communities. Second, our sample covers 17 of the group's roughly 30 active admins and moderators, including 1 who had since stepped down. We reached thematic saturation after 8 interviews as already mentioned in the methodology. The remaining moderators were not unwilling in principle, but other commitments meant we could not schedule interviews with them within our timeline. All admins and moderators active during our study period were male, participants reported that female moderators existed in the past, and we contacted two but received no response. Our findings thus reflect the perspectives of the group's current, all-male moderation team.

\section{Conclusion}

In this study, we examined why volunteer admins and moderators sustain \textit{HigherStudyAbroad}, a large, real-name Facebook group supporting Bangladeshi students pursuing higher study abroad. Across 17 interviews, we found altruism functioning not as motivation alone but as the organizing logic behind moderator identity, labor, and legitimacy, sustaining largely invisible labor and grounding the group's standing against paid consultancies. Participants expect this to hold even as AI absorbs routine questions, with the group narrowing toward the lived judgment no model can offer. Grounded in a single case, these findings add a high-stakes, real-name, Global South setting to CSCW's understanding of unpaid moderation, alongside its existing body of work on pseudonymous, Western platforms.

\bibliographystyle{ACM-Reference-Format}
\bibliography{lit_ref}

@misc{iom2024internationalstudents,
  author       = {{International Organization for Migration}},
  title        = {International Students},
  year         = {2024},
  howpublished = {World Migration Report},
  url          = {https://worldmigrationreport.iom.int/what-we-do/world-migration-report-2024/chapter-2/international-students},
  note         = {Accessed: 2026-08-29}
}

@article{simonds2023understanding,
  author  = {Simonds, Anne Niroshini and Abd Hamid, Junainah and Khatibi, Ali and Azam, S. M. Ferdous},
  title   = {Understanding International Student Mobility from a Developing Nation's Perspective},
  journal = {Russian Law Journal},
  year    = {2023},
  volume  = {11},
  number  = {7s},
  pages   = {125--144},
  url     = {https://www.russianlawjournal.org/index.php/journal/article/view/1085}
}

@techreport{unesco2026trends,
  author      = {{UNESCO International Institute for Higher Education in Latin America and the Caribbean}},
  title       = {Higher Education Global Trends Report: Towards Inclusive, Equitable and Quality Higher Education in an Internationally Mobile Landscape},
  institution = {UNESCO IESALC},
  year        = {2026},
  month       = may,
  url         = {https://www.iesalc.unesco.org/en/publication/higher-education-global-trends-report},
  urldate     = {2026-09-06}
}

@article{robinsonpant2018recruitment,
  author  = {Robinson-Pant, Anna and Magyar, Anna},
  title   = {The Recruitment Agent in Internationalized Higher Education: Commercial Broker and Cultural Mediator},
  journal = {Journal of Studies in International Education},
  year    = {2018},
  volume  = {22},
  number  = {3},
  pages   = {225--241},
  doi     = {10.1177/1028315318762485}
}

@article{xiang2014,
author = {Biao Xiang and Johan Lindquist},
title ={Migration Infrastructure},

journal = {International Migration Review},
volume = {48},
number = {1\_suppl},
pages = {122-148},
year = {2018},
doi = {10.1111/imre.12141}
}

@misc{unesco2026outbound,
  author       = {{UNESCO Institute for Statistics}},
  title        = {Total Outbound Internationally Mobile Tertiary Students Studying Abroad, All Countries, Both Sexes (UIS Estimate)},
  year         = {2026},
  howpublished = {\url{https://databrowser.uis.unesco.org/browser/EDUCATION/UIS-EducationOPRI/int-stud/outbound}},
  note         = {Data for Bangladesh; accessed 2026-09-08},
  organization = {UNESCO Institute for Statistics}
}

@incollection{rashid2024social,
  author    = {Rashid, Syeda Rozana},
  title     = {Social Backgrounds, Capitals and Risks in Youth Migration from Bangladesh},
  booktitle = {South Asia Migration Report 2024: Remittances, Resilience and Rehabilitation},
  editor    = {Rajan, S. Irudaya},
  year      = {2024},
  publisher = {Routledge India},
  address   = {London},
  doi       = {10.4324/9781003521228-7}
}

@article{ahmed2025braindrain,
  author  = {Ahmed, Rashed},
  title   = {Brain Drain: Out of Bound Mobility of International Students in {B}angladesh},
  journal = {Journal of Digital Learning and Distance Education},
  year    = {2025},
  volume  = {3},
  number  = {10},
  pages   = {1319--1326},
  doi     = {10.56778/jdlde.v3i10.454},
}

@article{islam2025migration,
  author    = {Md. Ariful Islam},
  title     = {Migration and Remittance in Bangladesh: Dynamics, Economic Impact, and Policy Implications},
  journal   = {Journal of Scientific Reports},
  volume    = {10},
  number    = {1},
  pages     = {144--167},
  year      = {2025},
  publisher = {IJSAB International},
  doi       = {10.58970/JSR.1121}
}

@article{RobinsonPant,
  author    = {Anna Robinson-Pant and Anna Magyar},
  title     = {The Recruitment Agent in Internationalized Higher Education: Commercial Broker and Cultural Mediator},
  journal   = {Journal of Studies in International Education},
  volume    = {22},
  number    = {3},
  pages     = {225--241},
  year      = {2018},
  doi       = {10.1177/1028315318762485}
}

@inproceedings{wohn2019,
author = {Wohn, Donghee Yvette},
title = {Volunteer Moderators in Twitch Micro Communities: How They Get Involved, the Roles They Play, and the Emotional Labor They Experience},
year = {2019},
isbn = {9781450359702},
publisher = {Association for Computing Machinery},
address = {New York, NY, USA},
url = {https://doi.org/10.1145/3290605.3300390},
doi = {10.1145/3290605.3300390},
booktitle = {Proceedings of the 2019 CHI Conference on Human Factors in Computing Systems},
pages = {1–13},
numpages = {13},
location = {Glasgow, Scotland Uk},
series = {CHI '19}
}

@inproceedings{dosono2019moderation,
author = {Dosono, Bryan and Semaan, Bryan},
title = {Moderation Practices as Emotional Labor in Sustaining Online Communities: The Case of AAPI Identity Work on Reddit},
year = {2019},
isbn = {9781450359702},
publisher = {Association for Computing Machinery},
address = {New York, NY, USA},
url = {https://doi.org/10.1145/3290605.3300372},
doi = {10.1145/3290605.3300372},
booktitle = {Proceedings of the 2019 CHI Conference on Human Factors in Computing Systems},
pages = {1–13},
numpages = {13},
location = {Glasgow, Scotland Uk},
series = {CHI '19}
}

@inproceedings{menking2015,
author = {Menking, Amanda and Erickson, Ingrid},
title = {The Heart Work of Wikipedia: Gendered, Emotional Labor in the World's Largest Online Encyclopedia},
year = {2015},
isbn = {9781450331456},
publisher = {Association for Computing Machinery},
address = {New York, NY, USA},
url = {https://doi.org/10.1145/2702123.2702514},
doi = {10.1145/2702123.2702514},
booktitle = {Proceedings of the 33rd Annual ACM Conference on Human Factors in Computing Systems},
pages = {207–210},
numpages = {4},
location = {Seoul, Republic of Korea},
series = {CHI '15}
}

@inproceedings{caiwohn2022,
author = {Cai, Jie and Wohn, Donghee Yvette},
title = {Coordination and Collaboration: How do Volunteer Moderators Work as a Team in Live Streaming Communities?},
year = {2022},
isbn = {9781450391573},
publisher = {Association for Computing Machinery},
address = {New York, NY, USA},
url = {https://doi.org/10.1145/3491102.3517628},
doi = {10.1145/3491102.3517628},
booktitle = {Proceedings of the 2022 CHI Conference on Human Factors in Computing Systems},
articleno = {300},
numpages = {14},
location = {New Orleans, LA, USA},
series = {CHI '22}
}

@article{gibson2023,
author = {Anna D. Gibson},
title ={What Teams Do: Exploring Volunteer Content Moderation Team Labor on Facebook},
journal = {Social Media + Society},
volume = {9},
number = {3},
pages = {20563051231186109},
year = {2023},
doi = {10.1177/20563051231186109},


}

@article{matias2019civic,
    author = {J. Nathan Matias},
    title = {The Civic Labor of Volunteer Moderators Online},
    journal = {Social Media + Society},
    volume = {5},
    number = {2},
    pages = {2056305119836778},
    year = {2019},
    doi = {10.1177/2056305119836778}
}

@article{seering2020,
author = {Seering, Joseph},
title = {Reconsidering Self-Moderation: the Role of Research in Supporting Community-Based Models for Online Content Moderation},
year = {2020},
issue_date = {October 2020},
publisher = {Association for Computing Machinery},
address = {New York, NY, USA},
volume = {4},
number = {CSCW2},
url = {https://doi.org/10.1145/3415178},
doi = {10.1145/3415178},
journal = {Proc. ACM Hum.-Comput. Interact.},
month = oct,
articleno = {107},
numpages = {28}
}

@incollection{schwartz1977,
title = {Normative Influences on Altruism},
editor = {Leonard Berkowitz},
series= {Advances in Experimental Social Psychology},
publisher = {Academic Press},
volume = {10},
pages = {221-279},
year = {1977},
issn = {0065-2601},
doi = {https://doi.org/10.1016/S0065-2601(08)60358-5},
url = {https://www.sciencedirect.com/science/article/pii/S0065260108603585},
author = {Shalom H. Schwartz}
}

@article{kimReceiversGiversUnderstanding2022,
  title = {From {{Receivers}} to {{Givers}}: {{Understanding Practice}} of {{Reciprocity}} in an {{Online Support Community}}},
  shorttitle = {From {{Receivers}} to {{Givers}}},
  author = {Kim, Chelsea and Wang, Hao-Chuan},
  year = 2022,
  month = mar,
  journal = {Proceedings of the ACM on Human-Computer Interaction},
  volume = {6},
  number = {CSCW1},
  pages = {1--17},
  issn = {2573-0142},
  doi = {10.1145/3512938},
  urldate = {2026-08-06},
  langid = {english}
}

@article{andreoni1990,
    author = {Andreoni, James},
    title = {Impure Altruism and Donations to Public Goods: A Theory of Warm-Glow Giving},
    journal = {The Economic Journal},
    volume = {100},
    number = {401},
    pages = {464-477},
    year = {1990},
    month = {06},
    issn = {0013-0133},
    doi = {10.2307/2234133},

}

@article{clary1998,
  author  = {Clary, E. Gil and Snyder, Mark and Ridge, Robert D. and Copeland, John and Stukas, Arthur A. and Haugen, Julie and Miene, Peter},
  title   = {Understanding and Assessing the Motivations of Volunteers: A Functional Approach},
  journal = {Journal of Personality and Social Psychology},
  year    = {1998},
  volume  = {74},
  number  = {6},
  pages   = {1516--1530},
  doi     = {10.1037/0022-3514.74.6.1516}
}

@article{clary1986,
 ISSN = {00093920, 14678624},
 URL = {http://www.jstor.org/stable/1130415},
 author = {E. Gil Clary and Jude Miller},
 journal = {Child Development},
 number = {6},
 pages = {1358--1369},
 publisher = {[Wiley, Society for Research in Child Development]},
 title = {Socialization and Situational Influences on Sustained Altruism},
 urldate = {2026-08-29},
 volume = {57},
 year = {1986}
}

@article{penner2005,
  author  = {Penner, Louis A. and Dovidio, John F. and Piliavin, Jane A. and Schroeder, David A.},
  title   = {Prosocial Behavior: Multilevel Perspectives},
  journal = {Annual Review of Psychology},
  year    = {2005},
  volume  = {56},
  pages   = {365--392},
  doi     = {10.1146/annurev.psych.56.091103.070141}
}

@article{yang2025takenforgranted,
  author    = {Ying Yang},
  title     = {Taken-for-granted Understandings in Access to International Higher Education: A Study of Chinese Agent-User Students' University Application Experiences},
  journal   = {Educational Review},
  volume    = {78},
  number    = {2},
  pages     = {203--222},
  year      = {2026},
  publisher = {Routledge},
  doi       = {10.1080/00131911.2025.2462636}
}

@article{salem2025perceptions,
  author    = {Sultan Salem and Samah Ali Mohsen Mofreh and Geethaletchmi Ponniah},
  title     = {Perceptions and Experiences with International Student Recruitment Agents: From the Students' Perspective},
  journal   = {Cogent Business \& Management},
  volume    = {12},
  number    = {1},
  pages     = {2555584},
  year      = {2025},
  publisher = {Cogent OA},
  doi       = {10.1080/23311975.2025.2555584}
}

@article{star1999,
  author  = {Susan Leigh Star and Anselm Strauss},
  title   = {Layers of Silence, Arenas of Voice: The Ecology of Visible and Invisible Work},
  journal = {Computer Supported Cooperative Work (CSCW)},
  year    = {1999},
  volume  = {8},
  pages   = {9--30},
  doi     = {10.1023/A:1008651105359}
}

@inproceedings{Mwesigwa2024,
author = {Mwesigwa, Daniel and Cs\'{\i}kszentmih\'{a}lyi, Christopher},
title = {Air/time Travel: Rethinking Appropriation in Global HCI and Futures of Electronic Exchange},
year = {2024},
isbn = {9798400703300},
publisher = {Association for Computing Machinery},
address = {New York, NY, USA},
url = {https://doi.org/10.1145/3613904.3642590},
doi = {10.1145/3613904.3642590},
booktitle = {Proceedings of the 2024 CHI Conference on Human Factors in Computing Systems},
articleno = {556},
numpages = {21},
location = {Honolulu, HI, USA},
series = {CHI '24}
}

@incollection{appadurai2004,
  author    = {Appadurai, Arjun},
  title     = {The Capacity to Aspire: Culture and the Terms of Recognition},
  booktitle = {Culture and Public Action},
  editor    = {Rao, Vijayendra and Walton, Michael},
  year      = {2004},
  pages     = {59--84},
  publisher = {Stanford University Press},
  address   = {Stanford, CA},
  url={https://gsdrc.org/document-library/the-capacity-to-aspire-culture-and-the-terms-of-recognition/}
}

@book{miyazaki2004b,
  author    = {Miyazaki, Hirokazu},
  title     = {The Method of Hope: Anthropology, Philosophy, and Fijian Knowledge},
  year      = {2004},
  publisher = {Stanford University Press},
  address   = {Stanford, CA},
  url={https://books.google.com.bd/books?hl=en&lr=&id=EHAtAi0C4lgC&oi=fnd&pg=PR13&dq=The+Method+of+Hope:+Anthropology,+Philosophy,+and+Fijian+Knowledge.&ots=kAIPYhn-Xz&sig=NWgpl2FUngStSxL5KQMM5YHwBfc&redir_esc=y#v=onepage&q=The%20Method%20of%20Hope%3A%20Anthropology%2C%20Philosophy%2C%20and%20Fijian%20Knowledge.&f=false}
}

@article{sabie2022,
author = {Sabie, Dina and Ekmekcioglu, Cansu and Ahmed, Syed Ishtiaque},
title = {A Decade of International Migration Research in HCI: Overview, Challenges, Ethics, Impact, and Future Directions},
year = {2022},
issue_date = {August 2022},
publisher = {Association for Computing Machinery},
address = {New York, NY, USA},
volume = {29},
number = {4},
issn = {1073-0516},
url = {https://doi.org/10.1145/3490555},
doi = {10.1145/3490555},
journal = {ACM Trans. Comput.-Hum. Interact.},
month = mar,
articleno = {30},
numpages = {35}
}

@article{braun2006using,
  author = {Braun, Virginia and Clarke, Victoria},
  title = {Using Thematic Analysis in Psychology},
  journal = {Qualitative Research in Psychology},
  volume = {3},
  number = {2},
  pages = {77--101},
  year = {2006},
  publisher = {Routledge},
  doi = {10.1191/1478088706qp063oa}
}

@book{corbin2008basics,
  author    = {Corbin, Juliet and Strauss, Anselm},
  title     = {Basics of Qualitative Research: Techniques and Procedures for Developing Grounded Theory},
  edition   = {3rd},
  year      = {2008},
  publisher = {Sage Publications},
  address   = {Thousand Oaks, CA},
  isbn      = {978-1412906449},
  url={https://methods.sagepub.com/book/mono/basics-of-qualitative-research/toc#_}
}

@article{piliavin1990,
  author  = {Piliavin, Jane Allyn and Charng, Hong-Wen},
  title   = {Altruism: A Review of Recent Theory and Research},
  journal = {Annual Review of Sociology},
  year    = {1990},
  volume  = {16},
  pages   = {27--65},
  doi     = {10.1146/annurev.so.16.080190.000331}
}

@article{lou2023,
  author  = {Lou, Katherine and Floridi, Luciano},
  title   = {Online Altruism: What It Is and How It Differs from Other Kinds of Altruism},
  journal = {Minds and Machines},
  year    = {2023},
  volume  = {33},
  number  = {4},
  pages   = {641--666},
  doi     = {10.1007/s11023-023-09648-8}
}

@incollection{kollock1999economies,
  title     = {The Economies of Online Cooperation: Gifts and Public Goods in Cyberspace},
  author    = {Kollock, Peter},
  booktitle = {Communities in Cyberspace},
  editor    = {Smith, Marc A. and Kollock, Peter},
  pages     = {220--239},
  year      = {1999},
  publisher = {Routledge},
  address   = {London},
  url = {https://dlc.dlib.indiana.edu/dlcrest/api/core/bitstreams/ccf091b6-d28b-40c3-bf78-da948819fc55/content}
}

@article{ren2007,
author = { Yuqing Ren and Robert Kraut and Sara Kiesler},
title ={Applying Common Identity and Bond Theory to Design of Online Communities},

journal = {Organization Studies},
volume = {28},
number = {3},
pages = {377-408},
year = {2007},
doi = {10.1177/0170840607076007},
}

@article{ren2012,
    author = {Yuqing Ren and F. Maxwell Harper and Sara Drenner and Loren Terveen and Sara Kiesler and John Riedl and Robert E. Kraut},
    title = {Building Member Attachment in Online Communities: Applying Theories of Group Identity and Interpersonal Bonds},
    journal = {Management Information Systems Quarterly},
    volume = {36},
    number = {3},
    pages = {841--863},
    year = {2012},
    month = {09},
    issn = {0276-7783},
    doi = {10.2307/41703483}
}

@article{preece2009reader,
  title={The reader-to-leader framework: Motivating technology-mediated social participation},
  author={Preece, Jennifer and Shneiderman, Ben},
  journal={AIS transactions on human-computer interaction},
  volume={1},
  number={1},
  pages={13--32},
  year={2009},
  url={https://aisel.aisnet.org/thci/vol1/iss1/5/}
}

@article{strauss1988,
  author  = {Anselm Strauss},
  title   = {The Articulation of Project Work: An Organizational Process},
  journal = {The Sociological Quarterly},
  year    = {2016},
  volume  = {29},
  number  = {2},
  pages   = {163--178},
  doi     = {10.1111/j.1533-8525.1988.tb01249.x}
}

@article{suchman1995,
  author  = {Suchman, Lucy},
  title   = {Making Work Visible},
  journal = {Communications of the ACM},
  year    = {1995},
  volume  = {38},
  number  = {9},
  pages   = {56--64},
  publisher = {Association for Computing Machinery},
address = {New York, NY, USA},
doi     = {10.1145/223248.223263},
}

@article{toxtli2021,
  title={Quantifying the invisible labor in crowd work},
  author={Toxtli, Carlos and Suri, Siddharth and Savage, Saiph},
  journal={Proceedings of the ACM on human-computer interaction},
  volume={5},
  number={CSCW2},
  pages={1--26},
  year={2021},
  publisher={Association for Computing Machinery},
  address={New York, NY, USA}, 
  doi={https://doi.org/10.1145/3476060}
}

@article{geiger2021,
author = {Geiger, R. Stuart and Howard, Dorothy and Irani, Lilly},
title = {The Labor of Maintaining and Scaling Free and Open-Source Software Projects},
year = {2021},
issue_date = {April 2021},
publisher = {Association for Computing Machinery},
address = {New York, NY, USA},
volume = {5},
number = {CSCW1},
url = {https://doi.org/10.1145/3449249},
doi = {10.1145/3449249},
journal = {Proc. ACM Hum.-Comput. Interact.},
month = apr,
articleno = {175},
numpages = {28}
}

@article{schopke2022,
    author  = {Sch{\"o}pke-Gonzalez, Angela M. and Atreja, Shubham and Shin, Han Na and Ahmed, Najmin and Hemphill, Libby},
    title   = {Why Do Volunteer Content Moderators Quit? Burnout, Conflict, and Harmful Behaviors},
    journal = {New Media \& Society},
    volume  = {26},
    number  = {10},
    pages   = {5677--5701},
    year    = {2022},
    doi     = {10.1177/14614448221138529}
}

@article{kuo2023unsung,
author = {Kuo, Tina and Hernani, Alicia and Grossklags, Jens},
title = {The Unsung Heroes of Facebook Groups Moderation: A Case Study of Moderation Practices and Tools},
year = {2023},
issue_date = {April 2023},
publisher = {Association for Computing Machinery},
address = {New York, NY, USA},
volume = {7},
number = {CSCW1},
url = {https://doi.org/10.1145/3579530},
doi = {10.1145/3579530},
journal = {Proc. ACM Hum.-Comput. Interact.},
month = apr,
articleno = {97},
numpages = {38}
}

@article{rubenstein2015,
author = {Rubenstein, Ellen L.},
title = {"They are always there for me": The convergence of social support and information in an online breast cancer community},
year = {2014},
issue_date = {July 2015},
publisher = {John Wiley \& Sons, Inc.},
address = {USA},
volume = {66},
number = {7},
issn = {2330-1635},
url = {https://doi.org/10.1002/asi.23263},
doi = {10.1002/asi.23263},
journal = {J. Assoc. Inf. Sci. Technol.},
month = jul,
pages = {1418–1430},
numpages = {13}
}

@book{titmuss1970,
   title={The gift relationship: From human blood to social policy},
  author={Titmuss, Richard M},
  year={2018},
  publisher={Policy Press},
  url=
  {https://books.google.com.bd/books?hl=en&lr=&id=_yxzDwAAQBAJ&oi=fnd&pg=PR1&dq=The+Gift+Relationship:+From+Human+Blood+to+Social+Policy&ots=Njcoif5hGk&sig=VS6osJMCljvshnIWdWvBXWs8eFQ&redir_esc=y}
    
}

@article{frey1997,
  author  = {Frey, Bruno S. and Oberholzer-Gee, Felix},
  title   = {The Cost of Price Incentives: An Empirical Analysis of Motivation Crowding-Out},
  journal = {American Economic Review},
  year    = {1997},
  volume  = {87},
  number  = {4},
  pages   = {746--755},
  url     = {https://www.jstor.org/stable/2951373}
}

@inproceedings{semaan2017saving,
author = {Semaan, Bryan and Dosono, Bryan and Britton, Lauren M.},
title = {Impression Management in High Context Societies: 'Saving Face' with ICT},
year = {2017},
publisher = {Association for Computing Machinery},
address = {New York, NY, USA},
doi = {10.1145/2998181.2998222},
pages = {712–725},
numpages = {14},
location = {Portland, Oregon, USA},
series = {CSCW '17}
}

@article{terranova2000,
  author  = {Terranova, Tiziana},
  title   = {Free Labor: Producing Culture for the Digital Economy},
  journal = {Social Text},
  year    = {2000},
  volume  = {18},
  number  = {2},
  pages   = {33--58},
  doi     = {10.1215/01642472-18-2_63-33}
}

@book{fuchs2014,
  author    = {Fuchs, Christian},
  title     = {Digital Labour and Karl Marx},
  year      = {2014},
  publisher = {Routledge},
  address   = {New York, NY},
  isbn      = {978-0-415-71615-4},
  url={https://books.google.com.bd/books?hl=en&lr=&id=GI6FAgAAQBAJ&oi=fnd&pg=PP1&dq=Digital+Labour+and+Karl+Marx&ots=sRhdIxSdko&sig=DNhPeZpU8lws9oWppwUfeevZ3_w&redir_esc=y#v=onepage&q=Digital%20Labour%20and%20Karl%20Marx&f=false}
}

@article{fourcade2020,
  author  = {Fourcade, Marion and Kluttz, Daniel N},
  title   = {A {M}aussian Bargain: Accumulation by Gift in the Digital Economy},
  journal = {Big Data \& Society},
  year    = {2020},
 url={https://journals.sagepub.com/doi/full/10.1177/2053951719897092}
}

@article{dekker2014,
  author  = {Dekker, Rianne and Engbersen, Godfried},
  title   = {How Social Media Transform Migrant Networks and Facilitate Migration},
  journal = {Global Networks},
  year    = {2013},
  volume  = {14},
  number  = {4},
  pages   = {401--418},
  doi     = {10.1111/glob.12040}
}

@article{batista2025,
  author  = {Batista, Catia and Han, Daniel and Haushofer, Johannes and Khanna, Gaurav and McKenzie, David and Mobarak, Ahmed Mushfiq and Theoharides, Caroline and Yang, Dean},
  title   = {Brain Drain or Brain Gain? Effects of High-Skilled International Emigration on Origin Countries},
  journal = {Science},
  year    = {2025},
  doi={https://doi.org/10.1126/science.adr8861}
}

@article{beine2008,
  author  = {Beine, Michel and Docquier, Fr{\'e}d{\'e}ric and Rapoport, Hillel},
  title   = {Brain Drain and Human Capital Formation in Developing Countries: Winners and Losers},
  journal = {The Economic Journal},
  year    = {2008},
  doi={https://doi.org/10.1111/j.1468-0297.2008.02135.x}
}

@article{sahay2023,
  title={What role does “hope” play in ICT4D research?},
  author={Sahay, Sundeep},
  journal={The Information Society},
  volume={40},
  number={1},
  pages={54--64},
  year={2024},
  publisher={Taylor \& Francis}, 
  doi={https://doi.org/10.1080/01972243.2023.2279966}
}

@inproceedings{kumar2019aspirations,
author = {Kumar, Neha and Wong-Villacres, Marisol and Karusala, Naveena and Vishwanath, Aditya and Kumar, Arkadeep and Ismail, Azra},
title = {Aspirations-based design},
year = {2019},
isbn = {9781450361224},
publisher = {Association for Computing Machinery},
address = {New York, NY, USA},
url = {https://doi.org/10.1145/3287098.3287117},
doi = {10.1145/3287098.3287117},
booktitle = {Proceedings of the Tenth International Conference on Information and Communication Technologies and Development},
articleno = {2},
numpages = {11},
location = {Ahmedabad, India},
series = {ICTD '19}
}

@inproceedings{dell2016,
author = {Dell, Nicola and Kumar, Neha},
title = {The Ins and Outs of HCI for Development},
year = {2016},
isbn = {9781450333627},
publisher = {Association for Computing Machinery},
address = {New York, NY, USA},
url = {https://doi.org/10.1145/2858036.2858081},
doi = {10.1145/2858036.2858081},
booktitle = {Proceedings of the 2016 CHI Conference on Human Factors in Computing Systems},
pages = {2220–2232},
numpages = {13},
location = {San Jose, California, USA},
series = {CHI '16}
}

@article{safir2025,
author = {Safir, Abdullah Hasan and Mim, Nusrat Jahan and Haque, S M Taiabul and Ahmed, Syed Ishtiaque},
title = {Resilient Hope: Identity, Extensibility, and Freedom as Design and Policy Considerations in Social Computing},
year = {2025},
issue_date = {December 2025},
publisher = {Association for Computing Machinery},
address = {New York, NY, USA},
volume = {3},
number = {4},
url = {https://doi.org/10.1145/3764665},
doi = {10.1145/3764665},
journal = {ACM J. Comput. Sustain. Soc.},
month = nov,
articleno = {33},
numpages = {28}
}

@article{he2024,
  author  = {He, Qinglai and Hong, Yili and Raghu, TS},
  title   = {Platform Governance with Algorithm-Based Content Moderation: An Empirical Study on Reddit},
  journal = {Information Systems Research},
  year    = {2024},
  doi={https://doi.org/10.1287/isre.2021.0036}
}

@article{seering2019moderator,
  author  = {Seering, Joseph and Wang, Tony and Yoon, Jina and Kaufman, Geoff},
  title   = {Moderator Engagement and Community Development in the Age of Algorithms},
  journal = {New Media \& Society},
  volume  = {21},
  number  = {7},
  pages   = {1417--1443},
  year    = {2019},
  doi     = {10.1177/1461444818821316}
}

@article{nov2007,
author = {Nov, Oded},
title = {What motivates Wikipedians?},
year = {2007},
issue_date = {November 2007},
publisher = {Association for Computing Machinery},
address = {New York, NY, USA},
volume = {50},
number = {11},
issn = {0001-0782},
url = {https://doi.org/10.1145/1297797.1297798},
doi = {10.1145/1297797.1297798},
journal = {Commun. ACM},
month = nov,
pages = {60–64},
numpages = {5}
}

@inproceedings{bryant2005,
author = {Bryant, Susan L. and Forte, Andrea and Bruckman, Amy},
title = {Becoming Wikipedian: transformation of participation in a collaborative online encyclopedia},
year = {2005},
isbn = {1595932232},
publisher = {Association for Computing Machinery},
address = {New York, NY, USA},
url = {https://doi.org/10.1145/1099203.1099205},
doi = {10.1145/1099203.1099205},
booktitle = {Proceedings of the 2005 ACM International Conference on Supporting Group Work},
pages = {1–10},
numpages = {10},
location = {Sanibel Island, Florida, USA},
series = {GROUP '05}
}

@article{jhaver2019automod,
author = {Jhaver, Shagun and Birman, Iris and Gilbert, Eric and Bruckman, Amy},
title = {Human-Machine Collaboration for Content Regulation: The Case of Reddit Automoderator},
year = {2019},
issue_date = {October 2019},
publisher = {Association for Computing Machinery},
address = {New York, NY, USA},
volume = {26},
number = {5},
issn = {1073-0516},
url = {https://doi.org/10.1145/3338243},
doi = {10.1145/3338243},
journal = {ACM Trans. Comput.-Hum. Interact.},
month = jul,
articleno = {31},
numpages = {35}
}

@article{chandrasekharan2019crossmod,
author = {Chandrasekharan, Eshwar and Gandhi, Chaitrali and Mustelier, Matthew Wortley and Gilbert, Eric},
title = {Crossmod: A Cross-Community Learning-based System to Assist Reddit Moderators},
year = {2019},
issue_date = {November 2019},
publisher = {Association for Computing Machinery},
address = {New York, NY, USA},
volume = {3},
number = {CSCW},
url = {https://doi.org/10.1145/3359276},
doi = {10.1145/3359276},
journal = {Proc. ACM Hum.-Comput. Interact.},
month = nov,
articleno = {174},
numpages = {30}
}

@inproceedings{sotudeh2022mentsum,
    title = "{M}ent{S}um: A Resource for Exploring Summarization of Mental Health Online Posts",
    author = "Sotudeh, Sajad  and
      Goharian, Nazli  and
      Young, Zachary",
    booktitle = "Proceedings of the Thirteenth Language Resources and Evaluation Conference",
    month = jun,
    year = "2022",
    address = "Marseille, France",
    publisher = "European Language Resources Association",
    url = "https://aclanthology.org/2022.lrec-1.287/",
    pages = "2682--2692"
}

@inproceedings{diakopoulos2011quality,
author = {Park, Deokgun and Sachar, Simranjit and Diakopoulos, Nicholas and Elmqvist, Niklas},
title = {Supporting Comment Moderators in Identifying High Quality Online News Comments},
year = {2016},
isbn = {9781450333627},
publisher = {Association for Computing Machinery},
address = {New York, NY, USA},
url = {https://doi.org/10.1145/2858036.2858389},
doi = {10.1145/2858036.2858389},
booktitle = {Proceedings of the 2016 CHI Conference on Human Factors in Computing Systems},
pages = {1114–1125},
numpages = {12},
location = {San Jose, California, USA},
series = {CHI '16}
}

@ARTICLE{zhang2021obsolete,
  author={Zhang, Haoxiang and Wang, Shaowei and Chen, Tse-Hsun and Zou, Ying and Hassan, Ahmed E.},
  journal={IEEE Transactions on Software Engineering}, 
  title={An Empirical Study of Obsolete Answers on Stack Overflow}, 
  year={2021},
  volume={47},
  number={4},
  pages={850-862},
  doi={10.1109/TSE.2019.2906315}
  }

@misc{hsabdfb,
  author       = {Higher Study Abroad},
  title        = {Higher Study Abroad - Facebook Group},
  year         = {2026},
  howpublished = {Online},
  url          = {https://www.facebook.com/groups/146324928762530},
  note         = {Accessed: 2026-08-31}
}

@article{batson1981,
  title = {Is Empathic Emotion a Source of Altruistic Motivation?},
  author = {Batson, C. Daniel and Duncan, Bruce D. and Ackerman, Paula and Buckley, Terese and Birch, Kimberly},
  year = 1981,
  journal = {Journal of Personality and Social Psychology},
  volume = {40},
  number = {2},
  pages = {290--302},
  publisher = {American Psychological Association},
  address = {US},
  issn = {1939-1315(Electronic),0022-3514(Print)},
  doi = {10.1037/0022-3514.40.2.290}
}

@article{yao2021joinstaygo,
author = {Yao, Zheng and Yang, Diyi and Levine, John M. and Low, Carissa A. and Smith, Tenbroeck and Zhu, Haiyi and Kraut, Robert E.},
title = {Join, Stay or Go? A Closer Look at Members' Life Cycles in Online Health Communities},
year = {2021},
issue_date = {April 2021},
publisher = {Association for Computing Machinery},
address = {New York, NY, USA},
volume = {5},
number = {CSCW1},
url = {https://doi.org/10.1145/3449245},
doi = {10.1145/3449245},
journal = {Proc. ACM Hum.-Comput. Interact.},
month = apr,
articleno = {171},
numpages = {22}
}

@inproceedings{yang2017commitment,
author = {Yang, Diyi and Kraut, Robert and Levine, John M.},
title = {Commitment of Newcomers and Old-timers to Online Health Support Communities},
year = {2017},
isbn = {9781450346559},
publisher = {Association for Computing Machinery},
address = {New York, NY, USA},
url = {https://doi.org/10.1145/3025453.3026008},
doi = {10.1145/3025453.3026008},
booktitle = {Proceedings of the 2017 CHI Conference on Human Factors in Computing Systems},
pages = {6363–6375},
numpages = {13},
location = {Denver, Colorado, USA},
series = {CHI '17}
}

@inproceedings{kolla2024llmmod,
  author    = {Kolla, Mahi and Salunkhe, Siddharth and Chandrasekharan, Eshwar and Saha, Koustuv},
  title     = {LLM-Mod: Can Large Language Models Assist Content Moderation?},
  booktitle = {Extended Abstracts of the 2024 CHI Conference on Human Factors in Computing Systems},
  series    = {CHI EA '24},
  articleno = {217},
  numpages  = {8},
  year      = {2024},
  publisher = {Association for Computing Machinery},
  address   = {New York, NY, USA},
  doi       = {10.1145/3613905.3650828}
}

@article{huang2025legitimacy,
  author    = {Huang, Tao},
  title     = {Content moderation by LLM: from accuracy to legitimacy},
  journal   = {Artificial Intelligence Review},
  volume    = {58},
  pages     = {320},
  year      = {2025},
  doi       = {10.1007/s10462-025-11328-1},
  publisher = {Springer}
}

@article{gertz2023migration,
  author  = {Dedecek Gertz, Helena},
  title   = {Migration and Education on Social Media: What Migrants Discuss about Education in {F}acebook Groups},
  journal = {Frontiers in Sociology},
  volume  = {8},
  articleno = {1177411},
  year    = {2023},
  doi     = {10.3389/fsoc.2023.1177411}
}

@article{gouldner1960,
 ISSN = {00031224},
 URL = {http://www.jstor.org/stable/2092623},
 author = {Alvin W. Gouldner},
 journal = {American Sociological Review},
 number = {2},
 pages = {161--178},
 publisher = {[American Sociological Association, Sage Publications, Inc.]},
 title = {The Norm of Reciprocity: A Preliminary Statement},
 urldate = {2026-08-29},
 volume = {25},
 year = {1960}
}

@article{kumar2024watch,
  author  = {Kumar, Deepak and AbuHashem, Yousef Anees and Durumeric, Zakir},
  title   = {Watch Your Language: Investigating Content Moderation with Large Language Models},
  journal = {Proceedings of the International AAAI Conference on Web and Social Media},
  volume  = {18},
  number  = {1},
  pages   = {865--878},
  year    = {2024},
  doi     = {10.1609/icwsm.v18i1.31358}
}

@inproceedings{lingel2014city,
  author    = {Lingel, Jessica and Naaman, Mor and boyd, danah},
  title     = {City, Self, Network: Transnational Migrants and Online Identity Work},
  booktitle = {Proceedings of the 17th ACM Conference on Computer Supported Cooperative Work \& Social Computing},
  series    = {CSCW '14},
  pages     = {1502--1510},
  year      = {2014},
  publisher = {ACM},
  address   = {New York, NY, USA},
  location  = {Baltimore, MD, USA},
  doi       = {10.1145/2531602.2531693}
}

@article{ratto2023,
  author  = {Ratto, Matt and others},
  title   = {Reopening, Repetition and Resetting: {HCI} and the Method of Hope},
  journal = {Proceedings of the ACM on Human-Computer Interaction},
  year    = {2023},
  doi={https://doi.org/10.1145/3610205}
}

@article{gibson2023teams,
  title={What teams do: Exploring volunteer content moderation team labor on Facebook},
  author={Gibson, Anna D},
  journal={Social Media+ Society},
  volume={9},
  number={3},
  pages={20563051231186109},
  year={2023},
  publisher={SAGE Publications Sage UK: London, England},
  doi={10.1177/20563051231186109}
}

@article{beech2018migration,
  author  = {Beech, Suzanne E.},
  title   = {Adapting to Change in the Higher Education System: International Student Mobility as a Migration Industry},
  journal = {Journal of Ethnic and Migration Studies},
  year    = {2018},
  volume  = {44},
  number  = {4},
  pages   = {610--625},
  doi     = {10.1080/1369183X.2017.1315515}
}

@inproceedings{cai2022coordination,
  title={Coordination and collaboration: how do volunteer moderators work as a team in live streaming communities?},
  author={Cai, Jie and Wohn, Donghee Yvette},
  booktitle={Proceedings of the 2022 CHI Conference on Human Factors in Computing Systems},
  pages={1--14},
  year={2022},
  doi = {https://doi.org/10.1145/3491102.3517628}
}

@article{zamanFactorsShapingBangladeshi2023,
title = {Factors Shaping {{Bangladeshi}} Students' Migration Decision Using Push--Pull Theory: A Focus Group Study},
author = {Zaman, Noshin Tasnim and Sohel, Md. Salman and Obaidullah, {\relax Md}. and Hossen, Md. Sohrab and Rahman, Md. Toufiqur and Sifullah, Md. Khaled and Sarker, Md. Fouad Hossain},
  year = 2023,
  month = dec,
  journal = {SN Social Sciences},
  volume = {4},
  number = {1},
  pages = {4},
  doi = {10.1007/s43545-023-00797-2}
}

@article{fenghorta2021,
  author  = {Feng, Siyuan and Horta, Hugo},
  title   = {Brokers of international student mobility: The roles and processes of education agents in {China}},
  journal = {European Journal of Education},
  year    = {2021},
  volume  = {56},
  number  = {2},
  pages   = {248--264},
  doi     = {10.1111/ejed.12442},
  issn    = {0141-8211}
}

@article{jayadeva2020keepcalm,
  author  = {Jayadeva, Sazana},
  title   = {Keep calm and apply to {Germany}: How online communities mediate transnational student mobility from {India} to {Germany}},
  journal = {Journal of Ethnic and Migration Studies},
  year    = {2020},
  volume  = {46},
  number  = {11},
  pages   = {2240--2257},
  doi     = {10.1080/1369183X.2019.1643230}
}

\appendix

\end{document}